\documentclass[a4paper,fleqn,usenatbib]{mnras}
\usepackage[T1]{fontenc}
\usepackage{ae,aecompl}

\usepackage{graphicx}	
\usepackage[export]{adjustbox}
\usepackage{amsmath}	
\usepackage{amssymb}	
\usepackage{subfig}
\usepackage{fancyhdr}
\usepackage{url}
\usepackage{float}
\usepackage{multirow}
\usepackage{appendix}
\usepackage[utf8]{inputenc}
\usepackage{framed}
\usepackage{tcolorbox}
\usepackage{adjustbox}

\usepackage{hyperref}
\newcommand{\metre}{\ensuremath{\,\textrm{m}}}

\usepackage{esvect}
\usepackage{scalerel}
\usepackage{tikz}
\usetikzlibrary{svg.path}

\definecolor{orcidlogocol}{HTML}{A6CE39}
\tikzset{
  orcidlogo/.pic={
    \fill[orcidlogocol] svg{M256,128c0,70.7-57.3,128-128,128C57.3,256,0,198.7,0,128C0,57.3,57.3,0,128,0C198.7,0,256,57.3,256,128z};
    \fill[white] svg{M86.3,186.2H70.9V79.1h15.4v48.4V186.2z}
                 svg{M108.9,79.1h41.6c39.6,0,57,28.3,57,53.6c0,27.5-21.5,53.6-56.8,53.6h-41.8V79.1z M124.3,172.4h24.5c34.9,0,42.9-26.5,42.9-39.7c0-21.5-13.7-39.7-43.7-39.7h-23.7V172.4z}
                 svg{M88.7,56.8c0,5.5-4.5,10.1-10.1,10.1c-5.6,0-10.1-4.6-10.1-10.1c0-5.6,4.5-10.1,10.1-10.1C84.2,46.7,88.7,51.3,88.7,56.8z};
  }
}

\newcommand\orcidicon[1]{\href{https://orcid.org/#1}{\mbox{\scalerel*{
\begin{tikzpicture}[yscale=-1,transform shape]
\pic{orcidlogo};
\end{tikzpicture}
}{|}}}}

\usepackage{hyperref} 

\title[APOPHIS - Effects of the 2029 Earth's Encounter on the Surface and Nearby Dynamics]
{APOPHIS - Effects of the 2029 Earth's Encounter on the Surface and Nearby Dynamics}

\author[G. Valvano; O. C. Winter; R. Sfair; R. Machado Oliveira; G. Borderes-Motta; T. S. Moura;]
        {G. Valvano$^{1}$\thanks{E-mail:  giulia.valvano@unesp.br}\orcidicon{0000-0002-7905-1788}\,
        O. C. Winter$^{1}$\thanks{E-mail:  othon.winter@unesp.br}\orcidicon{0000-0002-4901-3289}\,
        R. Sfair$^{1}$\thanks{E-mail: rafael.sfair@unesp.br}\orcidicon{0000-0002-4939-013X}
        R. Machado Oliveira$^{1}$\thanks{E-mail: rai.machado@unesp.br }\orcidicon{0000-0002-6875-0508}\,
    \newauthor
        G. Borderes-Motta$^{2}$\thanks{E-mail: gabriel.borderes@uc3m.es}\orcidicon{0000-0002-4680-8414}\,
        T. S. Moura$^{1}$\thanks{E-mail: santos.moura@unesp.br}\orcidicon{0000-0002-3991-8738}\
\\
$^{1}$ Grupo de Din\^amica Orbital e Planetologia, S\~ao Paulo State University, UNESP, Guaratinguet\'{a}, CEP 12516-410, 
  S\~{a}o Paulo, Brazil\\
  $^{2}$ Bioengineering and Aerospace Engineering Department, Universidad Carlos III de Madrid, Leganés, 28911, Madrid, Spain}

\date{Accepted xxx. Received xxx; in original form xxx}

\pubyear{2021}

\begin{document}
\label{firstpage}
\pagerange{\pageref{firstpage}--\pageref{lastpage}}
\maketitle

\begin{abstract}
The 99942 Apophis close encounter with Earth in 2029 may provide information about asteroid's physical characteristics and measurements of Earth's effects on the asteroid surface. In this work, we analysed the surface and the nearby dynamics of Apophis. The possible effects of its 2029 encounter on the surface and environment vicinity are also analysed. We consider a 340 metres polyhedron with a uniform density (1.29 g$\cdot$cm$^{-3}$, 2.2 g$\cdot$cm$^{-3}$ and 3.5 g$\cdot$cm$^{-3}$). The slope angles are computed, as well their variation that arises during the close approach. Such variation reaches 4$^{\circ}$ when low densities are used in our simulations and reaches 2$^{\circ}$ when the density is high. The zero-velocity curves, the equilibrium points, and their topological classification are obtained. We found four external equilibrium points and two of them are linearly stable. We also perform numerical simulations of bodies orbiting the asteroid, taking into account the irregular gravitational field of Apophis and two extra scenarios of perturbations: the solar radiation pressure and the Earth's perturbation during the close approach. The radiation pressure plays an important role in the vicinity of the asteroid, only cm-sized particles survived for the time of integration. For densities of 2.2 g$\cdot$cm$^{-3}$ and 3.5 g$\cdot$cm$^{-3}$, a region of 5 cm radius particles survived for 30 years of the simulation, and for 1.29 g$\cdot$cm$^{-3}$, only particles with 15 cm of radius survived. The ejections and collisions are about 30-50 times larger when the close encounter effect is added, but around 56-59\% of particles still survive the encounter.

\end{abstract}

\begin{keywords}
Celestial mechanics -- Astrometry and celestial mechanics, minor planets, asteroids, general -- Planetary Systems, minor planets, asteroids -- Planetary Systems, planets and satellites: dynamical evolution and stability -- Planetary Systems
\end{keywords}

\section{Introduction}

Upon its discovery on 2004 June at Kitt Peak by R.A. Tucker, D.J. Tholen, and F. Bernardi \citep{gar2004, tucker2004, Smalley2004}, 99942 Apophis (originally designated as 2004 MN4) has its orbit constantly monitored since it was reported a high probability of collision with Earth of more than 2\% \citep{chesley_2005}. Although this possibility was later discharged \citep{Larsen2004}, other potential impact or approaches with Earth were reported in the following years, and Apophis has been considered as a potentially hazardous asteroid (PHA). The trajectory and the future close encounters parameters were improved with ephemeris derived from radar observations in 2005-2006 from Arecibo radiotelescope \citep{giorgini2008predicting}. With this data, they were able to reduce orbital uncertainties and infer a nominal distance for the 2029 approach of about 6 Earth radii, predict a distance of 0.34 au for the 2036 encounter, and find that an impact probability still remains \citep{giorgini2008predicting}. The next approach will be the closest encounter with the Earth and will occur on 13 April 2029. Apophis will pass at a distance near to the geostationary orbit and a distance about one-tenth the distance between the Earth and Moon.

The predictions for the pre-2029 orbit turn to be well-defined, however, for the post-2029 orbit, the predictions are not well determined due to the uncertainties caused by perturbations. The Yarkovsky effect is a considerable source for the orbital uncertainties of Apophis' orbit \citep{giorgini2008predicting, farnocchia2013yarkovsky, vokrouhlicky2015yarkovsky}. Thus considering the Yarkovsky effect and using astrometric data from 2004-2008, \citet{farnocchia2013yarkovsky} presented a new impact risk for Apophis. They infer an impact probability of about $\sim$10$^{-9}$ for the 2036 encounter and $\sim$10$^{-6}$ for 2068. 

Observations from the 2021 encounter provided measurements that improved the fit for the orbit of Apophis and eliminate the possibility of impact for the next 100 years. It led Apophis to be removed from the ESA’s asteroid Risk List after remaining on this list for almost 17 years \footnote{\href{https://cneos.jpl.nasa.gov/sentry/removed.html}{https://cneos.jpl.nasa.gov/sentry/removed.html}}.

The 2029 flyby may provide an opportunity to improve the 3D shape model, investigate possible changes on the spin state, reshaping and effects of the encounter on the surface. \citet{scheeres2005abrupt} predicted that the terrestrial torques caused by the encounter will change Apophis' spin state drastically, and consequently the Yarkovsky accelerations. Conversely, \citet{souchay2018changes} showed that the changes in the spin rate may be small, and the larger effects may occur in the obliquity and precession in longitude. The effects of a close approach with the Earth also could cause material landslides and migration as is showed in \citet{binzel2010earth}. However, the numerical simulations made for \citet{yu2014numerical} using soft-sphere code implementation predicted that the effects of the tidal pertubations on the Apophis' surface may be small, but could produce small landslides.

Therefore, the closest approach of 2029 with the Earth may provide measures through observations before, during, and after the encounter. The observational data from these observations may improve some physical characteristics, the understanding of the effects of the closest encounters with the Earth, validate models about the material, and other possibilities. In this work, we used the convex shape model from \citet{Pravec2014} to analyse the surface and the nearby dynamics of the asteroid Apophis considering the effects caused by the 2029 closest approach with the Earth. The paper is composed of the following sections. In the next section, we present the asteroid model used in this work, discussing its polyhedral shape model, general characteristics and the 2029 encounter configuration. Section \ref{surface} introduces the gravitational potential and geopotential considering the polyhedra method. We discuss physical features using the slope angle and its variation due to the Earth perturbation. In Section \ref{stability}, we explore the nearby environment of Apophis by calculating the zero-velocity curves and equilibrium points. We also present a set of numerical simulations of a disc of particles around Apophis considering the gravitational field and two additional scenarios of perturbations: the solar radiation pressure and, in Section \ref{instability}, the Earth perturbation on the 2029 trajectory. Finally, in the last section we provide our final comments.

\section{The asteroid model}
\label{asteroid_model}

Apophis was observed by the VLT (Very Large Telescope) and the data obtained from a polarimetric observing campaign in 2006 resulted in an approximate diameter of 270 metres for the asteroid \citep{cellino2006albedo}. In 2012 and 2013, \citet{Pravec2014} made a photometric observational campaign and discovered that Apophis is a tumbling asteroid. They computed the Apophis’ tumbling spin state as a retrograde rotational period of 27.38 $\pm$ 0.07 hours and a precession period of 263 $\pm$ 6 hours. 

The 2012-2013 Apophis' apparition also allowed the construction of a convex shape model. The Apophis' shape model is represented as an unscaled polyhedron with 1014 vertices and 2024 triangular faces \citep{Pravec2014}. \citet{BROZOVIC2018115} also provided a shape model for the asteroid Apophis, but they used radar observations from Goldstone and Arecibo to improve the previous model. This improved shape model is a 340 metres polyhedron with 2000 vertices and 3996 faces. 

From thermal infrared observations, \citet{muller2014thermal} reported an approximated size of 375 metres for (99942) Apophis. \citet{licandro2015canaricam} also provided an estimated size for Apophis using this technique, however, they combined their data with previous thermal observations. The resulting size for Apophis ranges from 380 to 393 metres.

We adopted the Apophis' diameter presented by \citet{BROZOVIC2018115}, 340 metres, and the shape model provided by \citet{Pravec2014}, hence the model by \citet{BROZOVIC2018115} is not publicly available. We used the rotational period of 27.38 hours with an obliquity of 180$^{\circ}$ to reproduce the Apophis' retrograde rotation \citep{Pravec2014} and do not consider the precession period as it is 10 times larger than the rotational period. 

\begin{figure}
\begin{center}
\subfloat{\includegraphics[trim = 0mm 5.5cm 0mm 0mm,width=1\columnwidth]{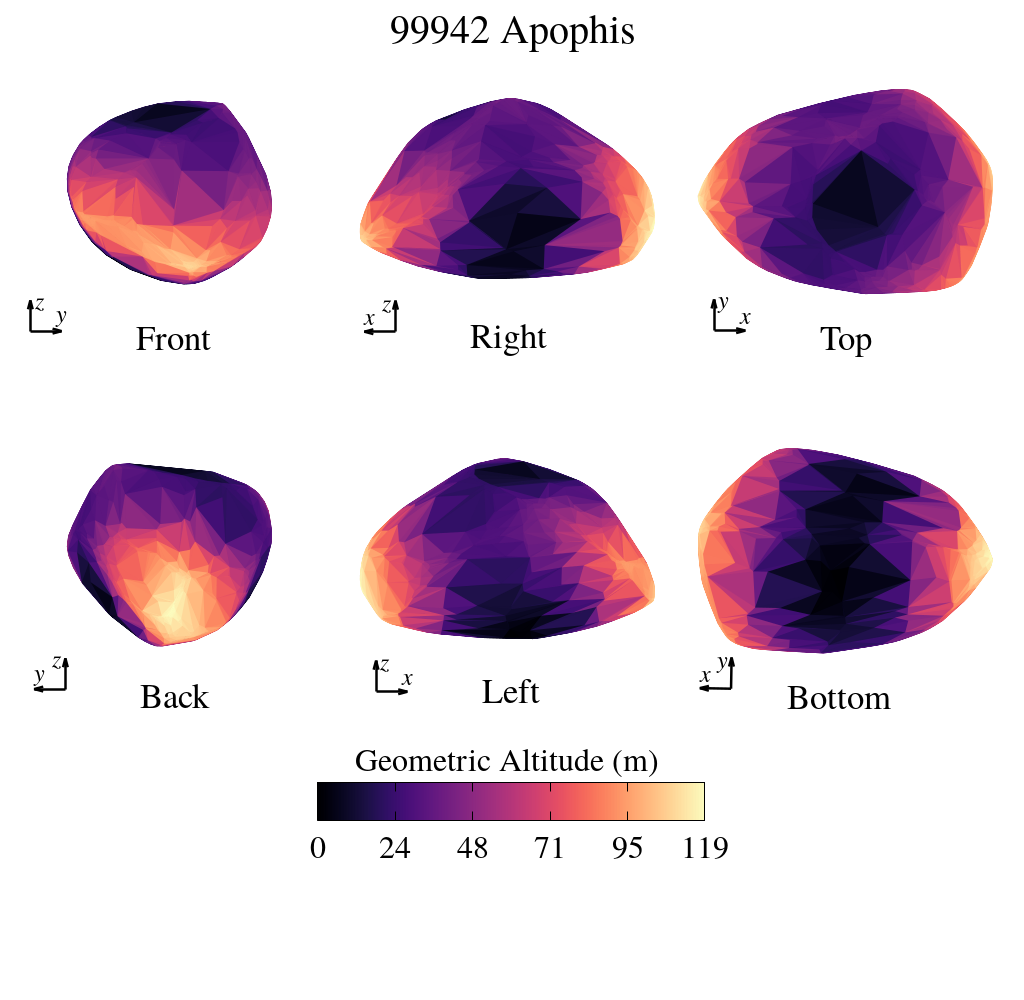}}
\end{center}
\caption{\label{fig:geom} Geometric altitude map computed across the
surface of Apophis under different views. The sea-level (the smaller distance between the geometric centre of the body and the barycentre among all triangular faces) is 130.54 metres.}
\end{figure}

The model for Apophis has an ellipsoid elongated format and tapered ends in the equatorial regions, as shown in its geometric map (Fig. \ref{fig:geom}). To measure the geometric altitude, we determine the distance between the geometric centre of the body and the barycentre of each triangular face of the polyhedron. Then, we identify the smaller distance between these points and set it as ``sea-level" \citep{Scheeres2016}. The sea-level is 130.54 metres and it is almost half the value of the maximum distance calculated between the geometric centre and the barycentre among all the faces.

Analysing the geometric map, we notice that the maximum values of the geometric altitude are on the equatorial regions, while the minimum values are on the poles, which means that these regions are at the sea-level. The Top and Bottom views in Fig. \ref{fig:geom} show the poles of the asteroid, and they shows that the north pole (Top view) has a concentrated region of the smaller altitudes on its middle. However, the south pole (Bottom view) presents a band of a larger distribution of small altitudes.

Apophis has a grain density in the range 3.4-3.6 g$\cdot$cm$^{-3}$ and a total porosity with range of 4-62\% because of its similar spectral characteristics with LL ordinary chondrite \citep{BINZEL2009480}. This results in an approximate bulk density of 1.29-3.46 g$\cdot$cm$^{-3}$. As the composition and some similarities suggest, Apophis could have a total porosity similar to the asteroid (25143) Itokawa. So \citet{BINZEL2009480} adopted Apophis' radius as 135 metres, density of 3.2 g$\cdot$cm$^{-3}$ and Itokawa's total porosity of 40\% to estimate Apophis' mass as 2.0$\times$10$^{10}$ kg. \citet{dachwald2007head} assumed a spherical radius of 160 metres and a bulk density of 2.72 g$\cdot$cm$^{-3}$ to determine the Apophis' mass as 4.67$\times$10$^{10}$ kg. \citet{muller2014thermal} also estimated a mass for Apophis, however they used a radius of 187.5 metres, density of 3.2 g$\cdot$cm$^{-3}$ and total porosity of 30-50\%, resulting in a total mass of 4.4-6.2$\times$10$^{10}$ kg.  Therefore, for this work, we assumed three bulk densities of 1.29 g$\cdot$cm$^{-3}$, 2.2  g$\cdot$cm$^{-3}$ and 3.5 g$\cdot$cm$^{-3}$.

Considering the density values adopted and the Apophis' size provided from \citet{BROZOVIC2018115}, we constructed three models for the asteroid, one for each density. As the mass is not a known parameter, we set the volume of the asteroid as the same of a sphere with an equivalent radius of 170 metres. So, we preserved the volume and the size and estimate the mass for each model according to the density. The volume adopted was 0.0205 km$^3$ and the three calculated masses are 2.64$\times$10$^{10}$ kg, 4.50$\times$10$^{10}$ kg and 7.16$\times$10$^{10}$ kg for the densities of 1.29 g$\cdot$cm$^{-3}$, 2.2 g$\cdot$cm$^{-3}$ and 3.5 g$\cdot$cm$^{-3}$, respectively.

With the three models, we defined a coordinate system with the origin at the object centre of mass and align the system axes to the axes of the principal moment of inertia. The $x$, $y$, and $z$ axes correspond, respectively, to the smallest, intermediate, and largest moments of inertia. This process was made using the algorithm presented by \citet{mir1996} and it was assumed that Apophis has a constant density and a uniform rotation about the largest moment of inertia. 

The values of the principal moment of inertia normalized by the mass are: 

\begin{equation}
I_{xx}/M = 559.23 \ \metre^{2},
\end{equation}
\begin{equation}
I_{yy}/M = 897.72 \ \metre^{2},
\end{equation}
\begin{equation}
I_{zz}/M = 969.39 \ \metre^{2}.
\end{equation}

From the principal moment of inertia, we calculated the second-order degree terms $C_{20}$ and $C_{22}$ of the gravity expansion. The coefficient $J_2$ ($-C_{20}$) represents how oblate Apophis is and $C_{22}$ how elongated. The values are \citep{macmillan1958theory, huscheeres2004}:

\begin{equation}
 C_{20}=-\frac{1}{2R_{n}^2}(2I_{zz}-I_{xx}-I_{yy})= -0.00837 \hspace{-0.4mm},
\label{eq:C20}  
\end{equation}
\begin{equation}
C_{22}=\frac{1}{4R_{n}^2}(I_{yy}-I_{xx})= 0.00294.
\label{eq: C22}  
\end{equation}
where the equivalent radius, $R_{equivalent} = $ 170 metres, was used for normalization.

Note that the coefficient $C_{20}$ is of the same order of magnitude as the coefficient $C_{22}$, but the absolute value of $C_{22}$ is smaller than $J_{2}$. We also modeled Apophis' shape as an ellipsoid with the parameters $a$, $b$ and $c$ representing the ellipsoid semi axes of the asteroid. The equivalent ellipsoid parameters for Apophis are:

\begin{equation}
a = 228.966631 \ \metre,
\end{equation}
\begin{equation}
b = 159.026432 \ \metre,
\end{equation}
\begin{equation}
c = 139.797960 \ \metre.
\end{equation}

Observe that $a$ is about 64\% larger than the parameter $c$, evidencing the elongated shape at the equator and flattened at the poles. So, considering the harmonic coefficients and the equivalent ellipsoid parameters, the shape of Apophis has a proportion of $5:3.5:3$ between the semi axes $a:b:c$, respectively.

\subsection{2029 Earth’s Encounter}

\begin{figure}
\begin{center}
\subfloat{\includegraphics[trim = 0mm 1.5cm 0mm 0mm,width=1\columnwidth]{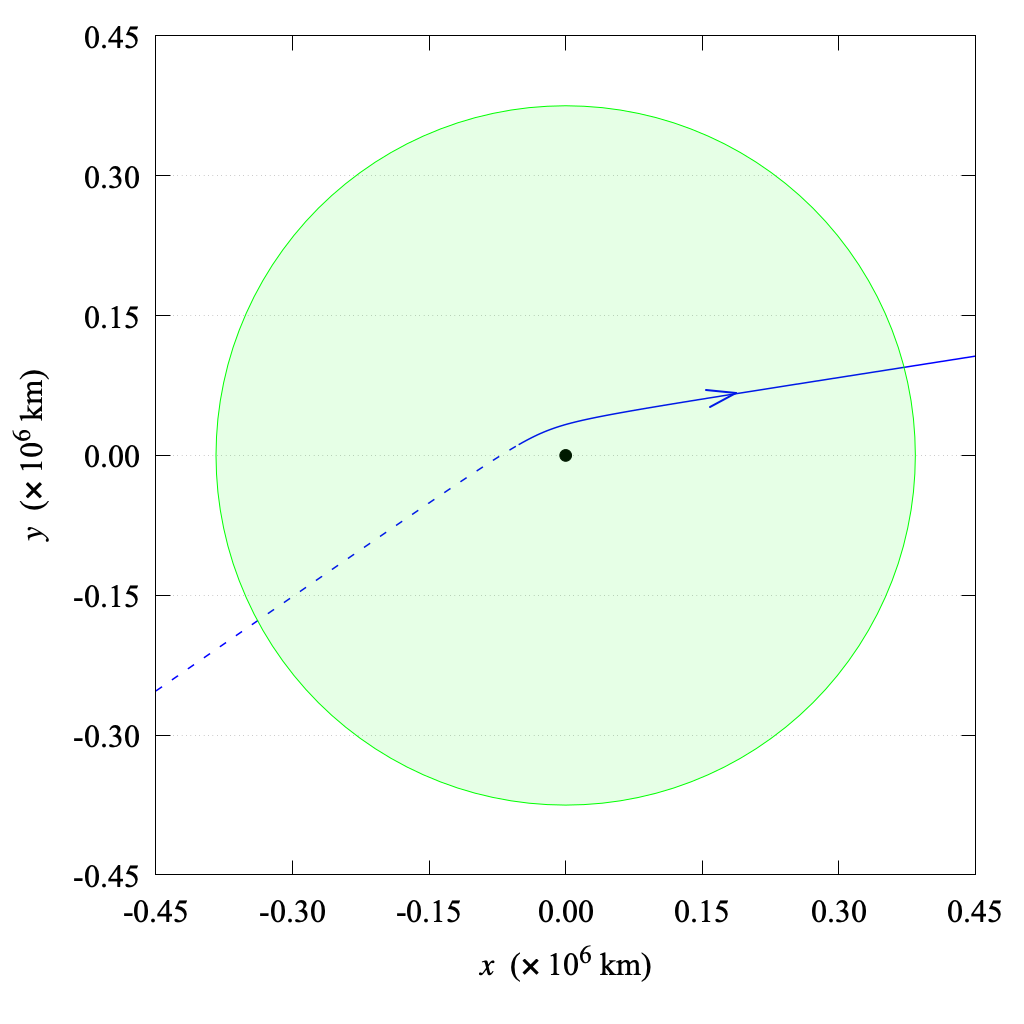}}
\end{center}
\caption{\label{fig:orbit_encounter} Apophis' trajectory of the encounter with Earth in 2029 in the $xoy$ plane. The black dot represents the Earth and, the green line and circle represent the Moon's orbit and the equatorial plane, respectively. The blue line represents the trajectory above the equatorial plane and the blue dashed line represents the trajectory below the equatorial plane. The arrow represents the movement direction of Apophis.}
\end{figure}

On April 13 2029, Apophis will have the closest approach with the Earth at an approximate distance of $\sim$38,000 km according to the JPL’s HORIZONS ephemerides\footnote{\href{https://ssd.jpl.nasa.gov/?horizons}{https://ssd.jpl.nasa.gov/?horizons}}. The trajectory is shown in Fig. \ref{fig:orbit_encounter}, where the black dot represents the Earth, the green circle and line illustrate the equatorial plane and Moon's orbit, respectively. The filled and the dashed blue lines represent the trajectory of the object above and below the equatorial plane, respectively, while the arrow illustrates the direction of the movement of Apophis. Note that Apophis approaches the Earth from left to right and it takes 34.30 hours to enter and leave the Moon's orbit. Apophis traverses the equatorial plane from the bottom-up and just crosses the equatorial plane near to the closest approach.

Since we do not know the exact orientation at the encounter, we investigate several hypothetical orientations for the asteroid Apophis at the moment of the encounter with the Earth in 2029 and see if this approach could change the surface characteristics of the body (section \ref{surface}). We also used the 2029 approach trajectory (Fig. \ref{fig:orbit_encounter}) to compute the effects on the dynamical nearby environment (section \ref{stability}). This analysis may provide insights to the 2029 observational campaign, and eventual space missions designed to study the asteroid.

\section{Effects on the Surface}
\label{surface}
The geopotential is an effective way to measure the relative energy on the body surface since it considers the gravitational and rotational potential \citep{Scheeres2012, Scheeres2016}. It is possible to relate the geopotential energy with the motion of a cohesionless particle by associating the quantity of energy necessary to move this particle across the body surface. 

Assuming a constant angular velocity vector $\pmb \omega$, the reference frame centred at the centre of mass and the axes aligned to the principal inertia axes for the asteroid Apophis, the expression for the geopotential is \citep{Scheeres2016}:
\begin{equation}
V(\pmb r) = -\frac{1}{2} \omega ^{2}(x^2+y^2) - U(\pmb r),
\label{eq:geopotencial}
\end{equation}
where $\pmb r$ is the position vector of a massless particle in the body-fixed frame and $U(\pmb r)$ the gravitation potential that is given by the method of polyhedra.

To model the object we used the polyhedra method, that computes the gravitational potential energy of an irregular body modeled by a uniform density polyhedron with a given number of faces and vertices. Then, the expression for the gravitational potential is \citep{wernerscheeres1996}:
\begin{equation}
U =\frac{G\rho}{2} \sum_{e\in edges}{\pmb r_e} \cdotp {\pmb E_e} \cdotp {\pmb r_e} \cdotp L_e -  \frac{G\rho}{2} \sum_{f\in faces}{\pmb r_f} \cdotp {\pmb F_f} \cdotp {\pmb r_f} \cdotp \omega_f,
 \label{eq: potencial}  
\end{equation}
where $\rho$ is the density of Apophis, $G$ is the gravitational constant; $\pmb r_f$ and $\pmb r_e$ represent, respectively, the position vectors from a point in the gravitational field to any point in the face $f$  and edge $e$ planes; $\pmb F_f$ and $\pmb E_e$ are the faces and edges tensors; $\omega_f$ and $L_e$ are the signed angle viewed from the field point and the integration factor, respectively.

From the geopotential, \citet{Scheeres2016} derive the slope angle using its gradient (see Section \ref{slope}), this measure considers the effects on the surface of the body, but we also can derive an approach for a particle in the nearby environment of the asteroid.

Since it is considered that Apophis has a constant angular velocity ($\pmb \omega$) about its axis of maximum moment of inertia, the movement of a particle orbiting the nearby environment is described by \citep{Scheeres2016}:
\begin{equation}
\ddot{\pmb r} +2{\pmb\omega} \times \dot{\pmb r} =-\frac{\partial V}{\partial \pmb {r}},
\label{eq: eqs do movimento}
\end{equation}
where $\pmb r$ is the position vector of the particle, $\dot{\pmb r}$ and $\ddot{\pmb r}$ are its velocity and acceleration vectors, respectively.

Equation \ref{eq: eqs do movimento} is time-invariant since the Apophis' spin-rate is assumed to be a constant value. Therefore this equation can be associated with a conserved quantity, $C_j$,  called ``Jacobi constant". This constant is defined by \citep {Scheeres2016}:
\begin{equation}
C_j = \frac{1}{2}v^2+V(\pmb r),
\label{eq: jacobi}
\end{equation}
where the magnitude of the velocity vector relative to the rotating frame of the Apophis is denoted by $v$.

\subsection{Slope}
\label{slope}

The slope angle is the supplement between the normal and total acceleration vectors \citep{Scheeres2012, Scheeres2016}. The slope is an angle that quantifies how inclined is a region on the Apophis' surface with respect to its local acceleration vector. Physically, slope assists in understanding the movement of free particles on the Apophis surface.

\begin{figure}
\begin{center}
\subfloat[$\rho$ = 1.29 g$\cdot$cm$^{-3}$]{\includegraphics*[trim = 0mm 4.5cm 0mm 0mm,
width=1\columnwidth]{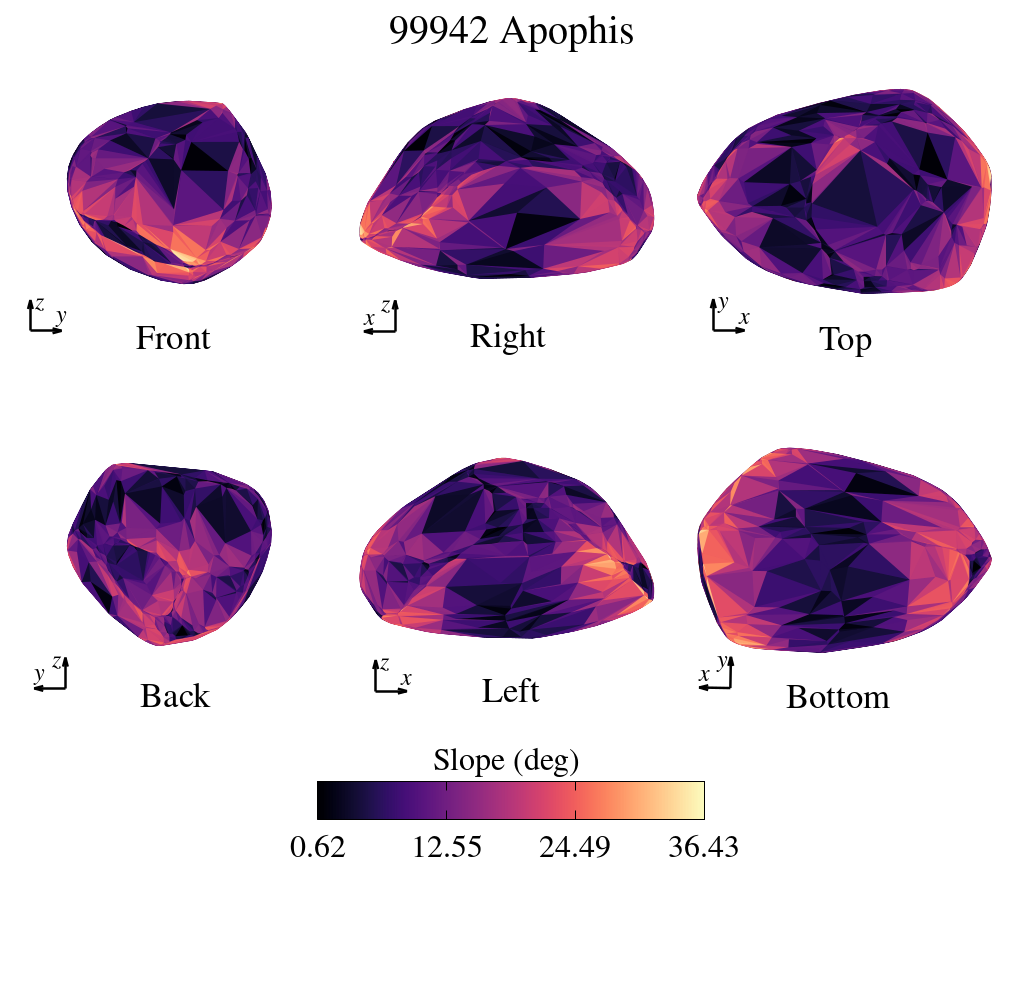}\label{fig:slope_129}}\\
\subfloat[$\rho$ = 3.5 g$\cdot$cm$^{-3}$]{\includegraphics*[trim = 0mm 4.5cm 0mm 0mm, width=1\columnwidth]{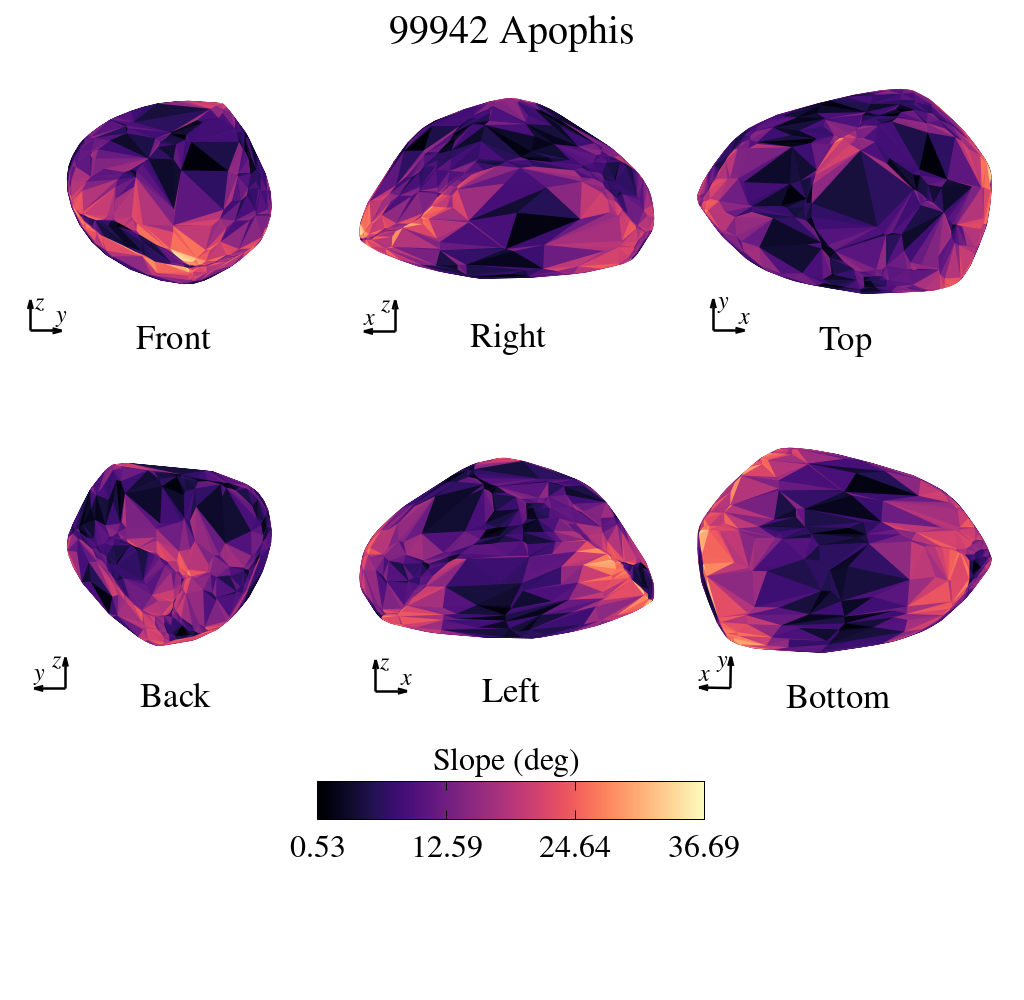}\label{fig:slope_35}}\\
\end{center}
\caption{\label{fig:slope} Slope angle maps across the surface of Apophis under different views. The letters (a) and (b) represent, respectively, the slope distribution considering the densities of 1.29 g$\cdot$cm$^{-3}$ and 3.5 g$\cdot$cm$^{-3}$.}
\end{figure}

The slope angle distribution across the Apophis' surface for the larger and smaller densities is shown in Fig. \ref{fig:slope}. The variation between the models is small and the slope distribution is almost identical. The slope variation amplitude goes from 35.81$^{\circ}$ for the smaller density, up to 36.16$^{\circ}$ for the larger density, a difference of only 0.35$^{\circ}$. 

For all three cases, the maximum slope angles are smaller than 37$^{\circ}$, and the minimum angles are larger than 0.5$^{\circ}$. The maximum values of slope occur just in the small regions on the equatorial extremities of Apophis (see front, left, and bottom views in Fig. \ref{fig:slope}). The minimum values of the slope are distributed preferentially at the central region of the body as we can see in the bottom view in Fig. \ref{fig:slope}, the south pole has a strip of minimum values is passing through the middle of the surface.

By definition, slope angles varies between 0$^{\circ}$ and 180$^{\circ}$. When it is larger than 90$^{\circ}$, the centrifugal force is larger than the gravitational force, causing cohesionless particles to escape from the body's surface. If the slope angle is smaller than 90$^{\circ}$, the movement of the cohesionless particles can be associated with the repose angle of a given material. The angle of repose for geological material is about 35-40 degrees \citep{lambe1969, apollo1974, al2018review}. 

As already referred, the maximum value of the slope considering the three values of density is about 37$^{\circ}$. So we can see two regimes, one for the lower limit value of the angle of repose, 35$^{\circ}$, and the other for the upper limit value, 40$^{\circ}$. For the lower limit, the flow of the cohesionless particles may occur, since we did not find a slope value higher than this limit.

\begin{figure}
\begin{center}
\subfloat{\includegraphics[trim = 0mm 4.5cm 0mm 0mm,width=1\columnwidth]{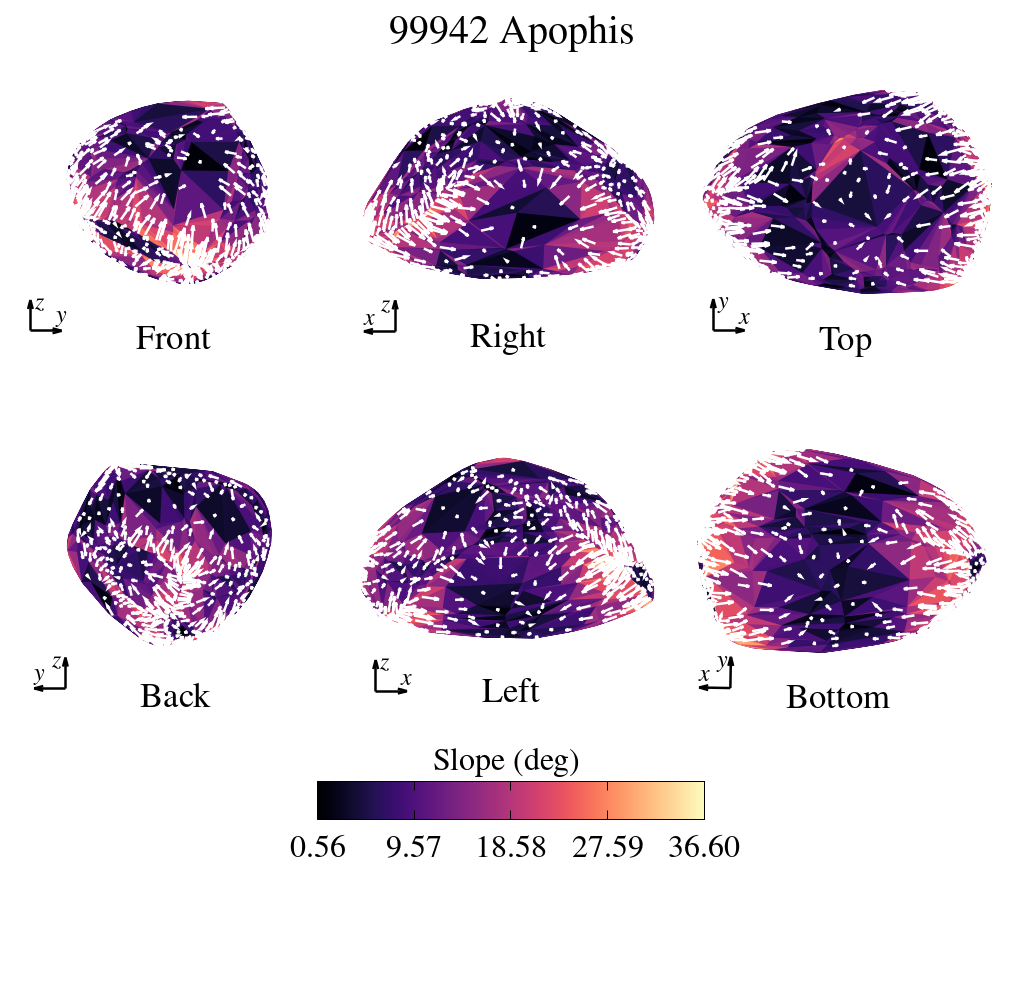}}
\end{center}
\caption{\label{fig:slope_setas} Slope angle map computed across the surface of Apophis with the directions of the local acceleration tangent vectors and under different views for the density of 2.2 g$\cdot$cm$^{-3}$.}
\end{figure}

To understand the possible flow of free particles on Apophis' surface, we compute the directions of the tangential component of the local acceleration vector \citep{Scheeres2012, Scheeres2016}. Figure \ref{fig:slope_setas} shows the tangential acceleration vectors considering a density of 2.2 g$\cdot$cm$^{-3}$. 

As expected, the vectors are pointing to the regions where the slope is smaller. The flow of loose material might occur from the small regions on the equatorial extremities to the middle of the Apophis' surface. The dark regions in Fig. \ref{fig:slope_setas} represent the near-zero slopes, thus are stable resting areas. Those regions are propitious regions to accumulate cohesionless particles.

\begin{figure}
\begin{center}
\subfloat[$\rho$ = 1.29 g$\cdot$cm$^{-3}$]{\includegraphics*[trim = 0mm 3cm 11cm 0mm,
width=0.9\columnwidth]{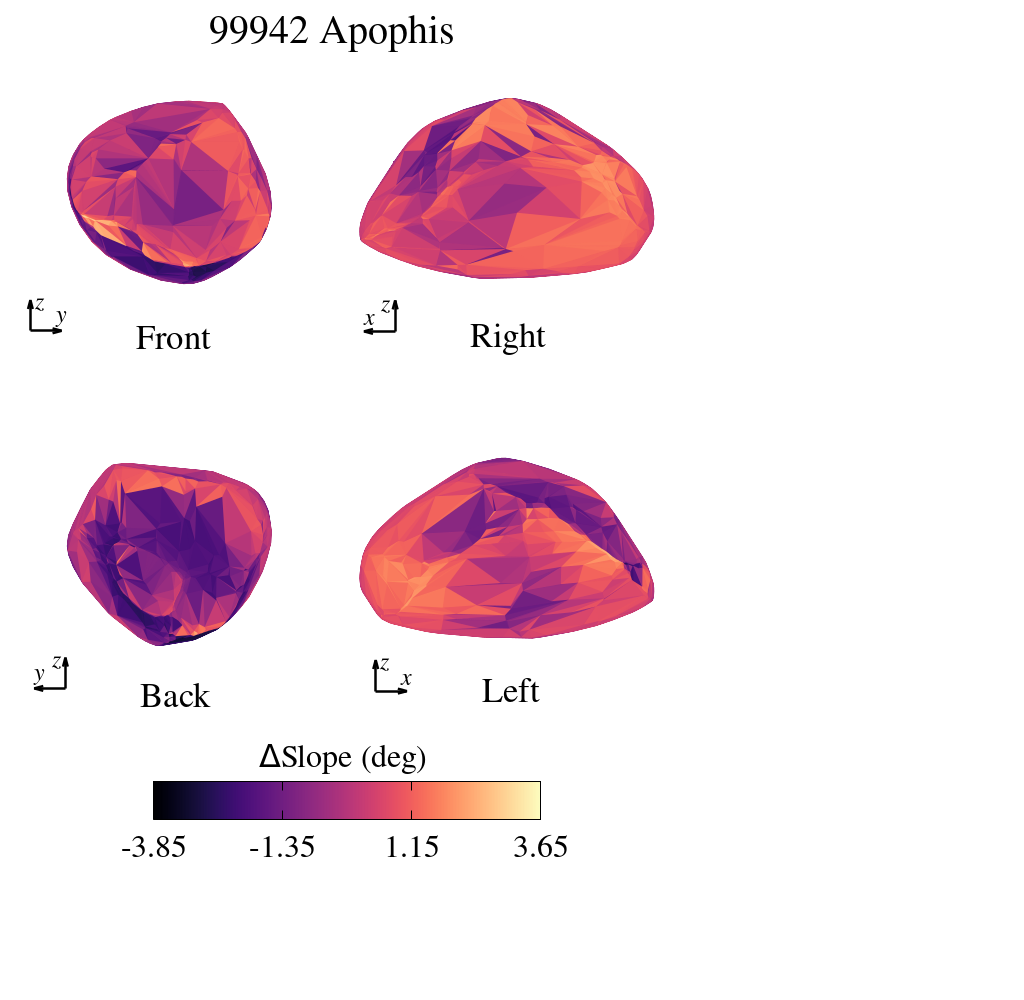}\label{fig:slope_129_terra}}\\
\subfloat[$\rho$ = 3.5 g$\cdot$cm$^{-3}$]{\includegraphics*[trim = 0mm 3cm 11cm 0mm, width=0.9\columnwidth]{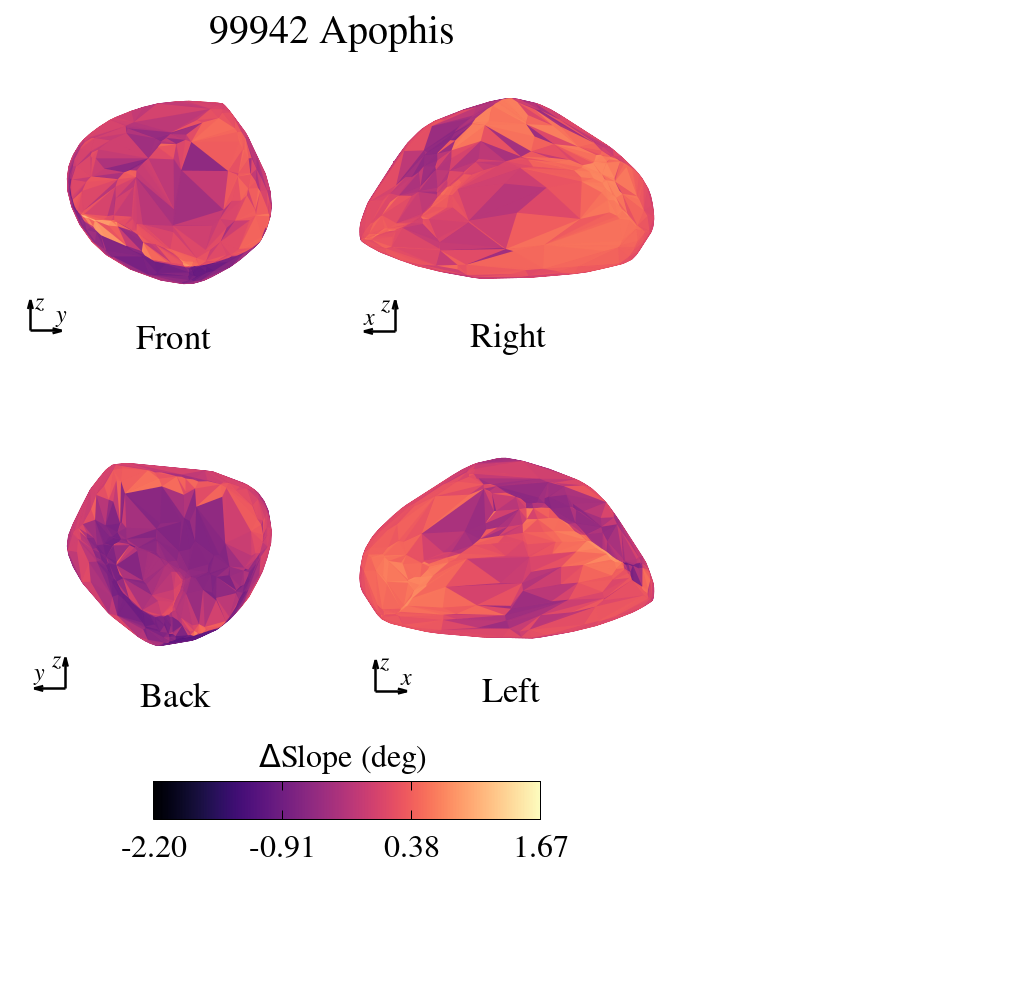}\label{fig:slope_35_terra}}\\
\end{center}
\caption{\label{fig:slope_terra} The $\Delta$slope angle maps across the surface of Apophis under different views and considering the Earth perturbation in 2029 encounter. The letters (a) and (b) represent, respectively, the slope distribution considering the densities of 1.29 g$\cdot$cm$^{-3}$ and 3.5 g$\cdot$cm$^{-3}$.}
\end{figure}

Apophis will have a close approach with the Earth in 2029 and will be affected by time-varying forces. Thus, to evaluate the strongest possible effect on its surface due to the Earth`s gravity, we calculated a $\Delta$slope that computes the slope variation caused due to the encounter at the closest distance of the whole approach.

In order to identify the most extreme condition that may occur in 2029, we assume a hypothetical configuration for the Apophis-Earth system. We set the Earth on the equatorial plane of Apophis and the distance between the centre of mass of Apophis and the Earth as the closest distance during the whole encounter, and  we calculate the slope on the Apophis' surface considering its geopotential including the Earth's gravitational force on the closest approach. Then, we rotate the shape model of Apophis all along the 360$^{\circ}$, so that each Apophis' orientation experiences the extreme approach condition.

The difference between the total acceleration vectors, with and without the Earth's gravity, leads to a slope that we call a delta slope. The delta slope is the variation of the slope angle produced by the Earth’s gravitational perturbation on the surface of Apophis at the extreme approach condition. The $\Delta$slope map was also made over the rotation of 360$^{\circ}$ of the Apophis’ shape model, in order to compute the $\Delta$slope on each Apophis' orientation.

Figure \ref{fig:slope_terra} shows the $\Delta$slope for the smaller and larger density of Apophis (see the complete animation of the $\Delta$slope angle maps in the videos of the complementary material). The top and bottom views are not shown since they are not directly pointed to the Earth at the closest approach, as we defined before. For the 1.29 g$\cdot$cm$^{-3}$ density model, the slope variation was smaller than 4$^{\circ}$, and for 3.5 g$\cdot$cm$^{-3}$ density model was about 2$^{\circ}$. This variation represents about 11\% of the regular slope angle for the smaller density model and 5.5\% for the larger density model.  

With the perturbation caused by the Earth, some slope angle values may exceed the angle of repose. \citet{ballouz2019surface} showed that a variation around 2$^{\circ}$ may sufficient to start a slow erosion process in regions with high-slope that experiment larger perturbations. However, the perturbations due to the 2029 encounter will not trigger drastic reshaping on Apophis' surface and shape. The numerical simulations provided by \citet{yu2014numerical} revealed that the effects caused by the Earth are small and just can cause local landslides.

\section{Effects on the Nearby Stability}
\label{stability}

In this section, we present an investigation of the environment around Apophis. The analysis of the variation of the Jacobi constant, through zero-velocity curves, allows us to identify where the movement is permitted or not. These curves delimit the external equilibrium points of the body, the location and the topological classification of these points are presented. A set of numerical simulations with a disc of particles is also presented in order to analyse the stability around the asteroid considering the perturbation of its own gravitational field and two additional perturbations: the solar radiation pressure and the 2029 flyby scenario.

\subsection{Equilibrium Points and Zero-Velocity Curves}
\label{points}

To understand the stability in the vicinity of Apophis, we calculated the zero-velocity curves and equilibrium points. The zero-velocity curves limit the movement of a particle, delimiting where its movement is allowed or not. This delimited movement depends on the Jacobi constant, C$_j$, and,  it arises from the inequality
\begin{equation}
C_j - V(\pmb r) \ge 0,
\label{eq: inequality}
\end{equation}
since the velocity term, $\frac{1}{2}v^2$, from equation \ref{eq: jacobi} shall be always positive. So, when $ C_j < V(\pmb r)$ there will be forbidden regions to the movement of the particle, considering that the inequality is infringed. In the case of the regions where $ C_j > V(\pmb r)$, there will be no a priori limitation to the movement of the particle, since the inequality is satisfied. When $ C_j = V(\pmb r)$ we have the zero-velocity curves that separate the regions between the permitted and prohibited movement.

\begin{figure}
\begin{center}
\includegraphics*[trim = 0mm 0cm 0mm 0cm,
width=1\columnwidth]{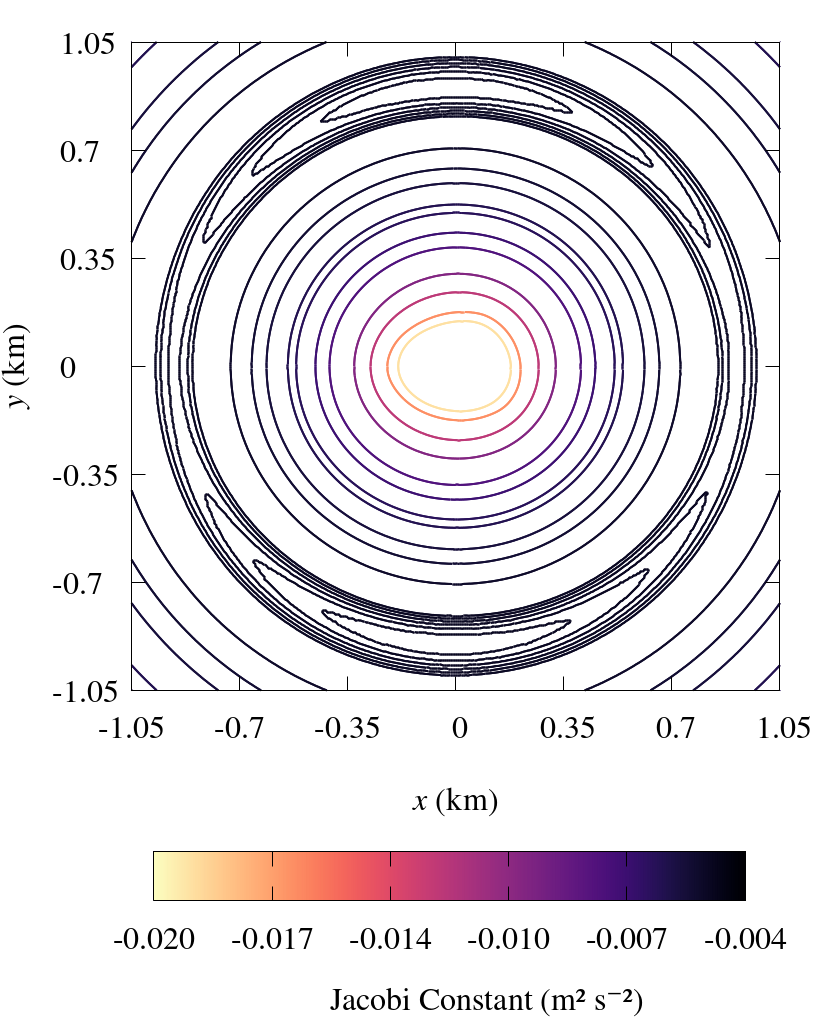}
\end{center}
\caption{\label{fig:cvz}  Zero-velocity curves contourplots in the $xoy$ plane considering the mean density of 2.2 g$\cdot$cm$^{-3}$. The colorbox represents the value of the Jacobi constant.}
\end{figure}

Fig. \ref{fig:cvz} illustrates the zero-velocity curves countourplots in the $xoy$ plane for the model with the mean density. The colour of the zero-velocity curves changes as the Jacobi constant value changes. At a certain value of C$_j$, the curves could form confined regions that will encompass an equilibrium point, similar to a ``banana” shape in Fig. \ref{fig:cvz}. The region between these confined curves also will include an equilibrium point.

The equilibrium points are critical points of geopotential where there is a balance between the gravitational and centrifugal forces. Since the resulting force is null at this points, the equilibrium points of the asteroid Apophis are calculated by the solution of:
\begin{equation}
\nabla V = 0,
\label{eq: eq_point}
\end{equation}
where the $\nabla V$ means the gradient of the geopotential.

In general, there is no fixed number of solutions for equation \ref{eq: eq_point}, since the solution depends on the body shape model, its density, and rotational period. However, we could estimate the number of equilibrium points just by looking at the zero-velocity curves of the body. The zero-velocity curves of Apophis show it has at least four equilibrium points, two of them inside the curves with a ``banana” shape  and two between the larger ``banana” curves near to the $y=0$ line (Fig. \ref{fig:cvz}).

The location of the equilibrium points also depends on the rotational period, density, and shape of the asteroid. Once defined the shape model and the spin, if we change its density, the location of the equilibrium points will change as well. If the density decreases, the location of the equilibrium points will be closer to the body due to the reduction of the gravitational force. Conversely, if the density is increased, the equilibrium points will be more distant from the body. 

\begin{figure}
\begin{center}
\includegraphics*[trim = 0mm 2.5cm 0mm 6cm,
width=1\columnwidth]{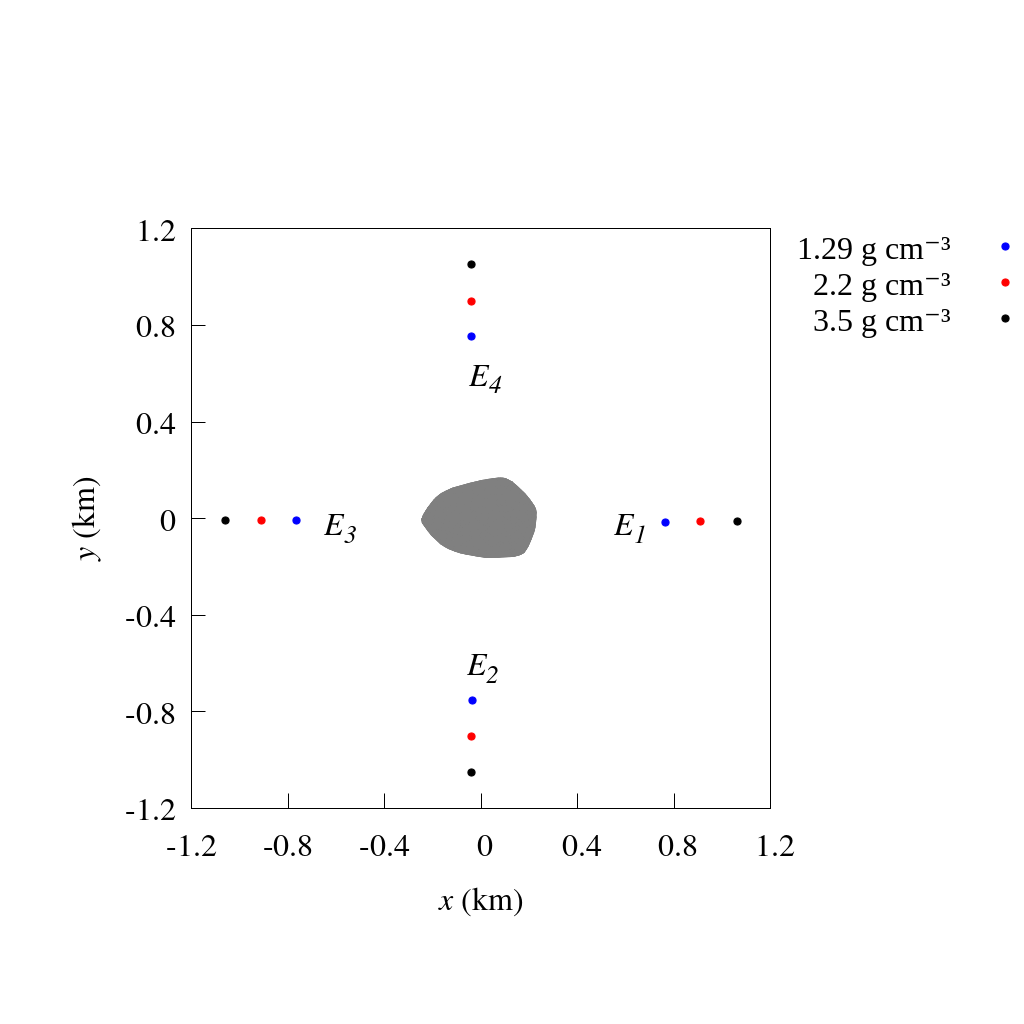}
\end{center}
\caption{\label{fig:pontos_eq}  Location of the equilibrium points in the $xoy$ plane for each density. The point's color represents the density of Apophis.}
\end{figure}

Considering the polyhedral model derived from \citet{Pravec2014} and the densities of 1.29 g$\cdot$cm$^{-3}$,  2.2 g$\cdot$cm$^{-3}$, and 3.5 g$\cdot$cm$^{-3}$, we found five equilibrium points for each density, being one in the centre of the body. The position of the equilibrium points are shown in table \ref{table:autovalores}. We note they are close to the equatorial plane, so Fig. \ref{fig:pontos_eq} shows the projection of the asteroid Apophis and the equilibrium points in the $xoy$ plane. 

Note that the number of equilibrium points does not change, differently from other systems \citet{jiang2018annihilation}. Observe that although the model with a larger density is almost three times larger than the model with a smaller density, the difference in the radial distance between the equilibrium points of the model with larger and smaller densities is about 300 metres.

\begin{table*}
\centering
\caption{Coordinates and characteristic time of the equilibrium points of 99942 Apophis considering the three densities (1.29 g$\cdot$cm$^{-3}$,  2.2 g$\cdot$cm$^{-3}$, and 3.5 g$\cdot$cm$^{-3}$) and their topological structures.}
\label{table:autovalores} 
\begin{tabular}{c|r|r|r|r|c}
\hline\hline
\multicolumn{1}{c}{{Point}} & \multicolumn{1}{c}{{$x$ (km)}} & \multicolumn{1}{c}{{$y$ (km)}} & \multicolumn{1}{c}{{$z$ (km)}} & \multicolumn{1}{c}{{Topological classification}} & \multicolumn{1}{c}{{ \begin{tabular}{@{}c@{}}Characteristic Time \\ (hours)\end{tabular}}}\\
\hline\hline
\multicolumn{6}{|c|}{$\rho$ = 1.29 g$\cdot$cm$^{-3}$}\\
\hline
    $E_1$ & 0.764033 & -0.014211 & -0.001470 & Saddle-Centre-Centre & \begin{tabular}{@{}c@{}}26.83192 \\ 26.97251\end{tabular}\\
     \hline
    $E_2$ & -0.038260& -0.752882 &  0.000598 & Centre-Centre-Centre & \begin{tabular}{@{}c@{}}27.28828 \\ 28.90402 \\ 88.39612\end{tabular}\\
     \hline
    $E_3$ & -0.765131& -0.007504 & -0.001544 & Saddle-Centre-Centre & \begin{tabular}{@{}c@{}}26.57328 \\ 26.90200\end{tabular}\\  
     \hline
    $E_4$ & -0.041584& 0.752713  & 0.000675 & Centre-Centre-Centre & \begin{tabular}{@{}c@{}}27.30720 \\ 28.79664 \\ 90.94898\end{tabular}\\ 
\hline \hline
\multicolumn{6}{c}{$\rho$ = 2.2 g$\cdot$cm$^{-3}$}\\
\hline
    $E_1$ & 0.910268  & -0.013482 & -0.001033 & Saddle-Centre-Centre & \begin{tabular}{@{}c@{}}26.98514 \\ 27.08051\end{tabular}\\
     \hline
    $E_2$ & -0.038718 & -0.900849 &  0.000424 & Centre-Centre-Centre & \begin{tabular}{@{}c@{}}27.31676 \\ 28.38482 \\ 107.46649 \end{tabular}\\
     \hline
    $E_3$ & -0.911036 & -0.007877 & -0.001078 & Saddle-Centre-Centre & \begin{tabular}{@{}c@{}}26.82109 \\ 27.04486\end{tabular}\\  
     \hline
    $E_4$ & -0.041482 & 0.900733  & 0.000470 & Centre-Centre-Centre & \begin{tabular}{@{}c@{}}27.32790 \\ 28.32686 \\ 110.03305\end{tabular}\\ 
\hline \hline
\multicolumn{6}{c}{$\rho$ = 3.5 g$\cdot$cm$^{-3}$}\\
\hline
    $E_1$ & 1.060747  & -0.012966 & -0.000760 & Saddle-Centre-Centre & \begin{tabular}{@{}c@{}}27.08273 \\ 27.15419\end{tabular}\\
     \hline
    $E_2$ & -0.039031 & -1.052625 & 0.000314 & Centre-Centre-Centre & \begin{tabular}{@{}c@{}}27.33415 \\ 28.09159 \\ 126.74930\end{tabular}\\
     \hline
    $E_3$ & -1.061310 & -0.008169 & -0.000789 & Saddle-Centre-Centre & \begin{tabular}{@{}c@{}}26.97445 \\ 27.13409\end{tabular}\\  
     \hline
    $E_4$ & -0.041388 & 1.052541  & 0.000343 & Centre-Centre-Centre & \begin{tabular}{@{}c@{}}27.34118 \\ 28.05680 \\ 129.32771\end{tabular}\\ 
\hline \hline
\end{tabular}
\end{table*}

Applying the linearization method to the equations of motion, we analysed the six eigenvalues of the characteristic equation for each point in order to identify their topological classification \citep{Jiang2014}. Table \ref{table:autovalores} shows the coordinates and the topological classification of each equilibrium point for each density of Apophis. The odd indexed points, $E_1$ and $E_3$, are classified as a Saddle-Center-Center, while the even points ($E_2$ and $E_4$) are Center-Center-Center, implying that they are linearly stable points. 

According to \citet{Jiang2014} and \citet{yu2012orbital}, when analysing the eigenvalues of the characteristic equation, we identified the existence of three families of periodic orbits in the tangent space of the equilibrium points $E_2$ and $E_4$, which have characteristic times or oscillation periods of approximately 27.3 hr, 28.9 hr and 88.4 hr for $E_2$, and 27.3 hr, 28.8 hr and 90.9 hr for $E_4$, when we consider the density of 1.29 g$\cdot$cm$^{-3}$. Only the period value of the third periodic orbit family underwent a significant change, with an increase of about 21\% and 43\%, respectively, for the Apophis intermediate and upper densities, when we analysed the $E_2$ equilibrium point eigenvalues. The same conclusion can be applied to point $E_4$. While for points $E_1$ and $E_3$, there are two families of periodic orbits, which have periods of oscillation around 26.8 hr and 26.9 for $E_1$, and 26.6 hr and 26.9 hr for $E_3$, and these values undergo small changes, less than 30 minutes, for other Apophis densities (Table \ref{table:autovalores}).

Note that despite of changing the density of Apophis, the topological classification of the equilibrium points $E_2$ and $E_4$ remains as linearly stable. Thus, we performed numerical simulations of a disc of particles encompassing the equilibrium points regions in order to identify possible stable zones around them.

\subsection{Stability Regions}
\label{regions}

\begin{figure}
\begin{center}
\includegraphics*[trim = 0cm 0cm 0cm 0mm,
width=1\columnwidth]{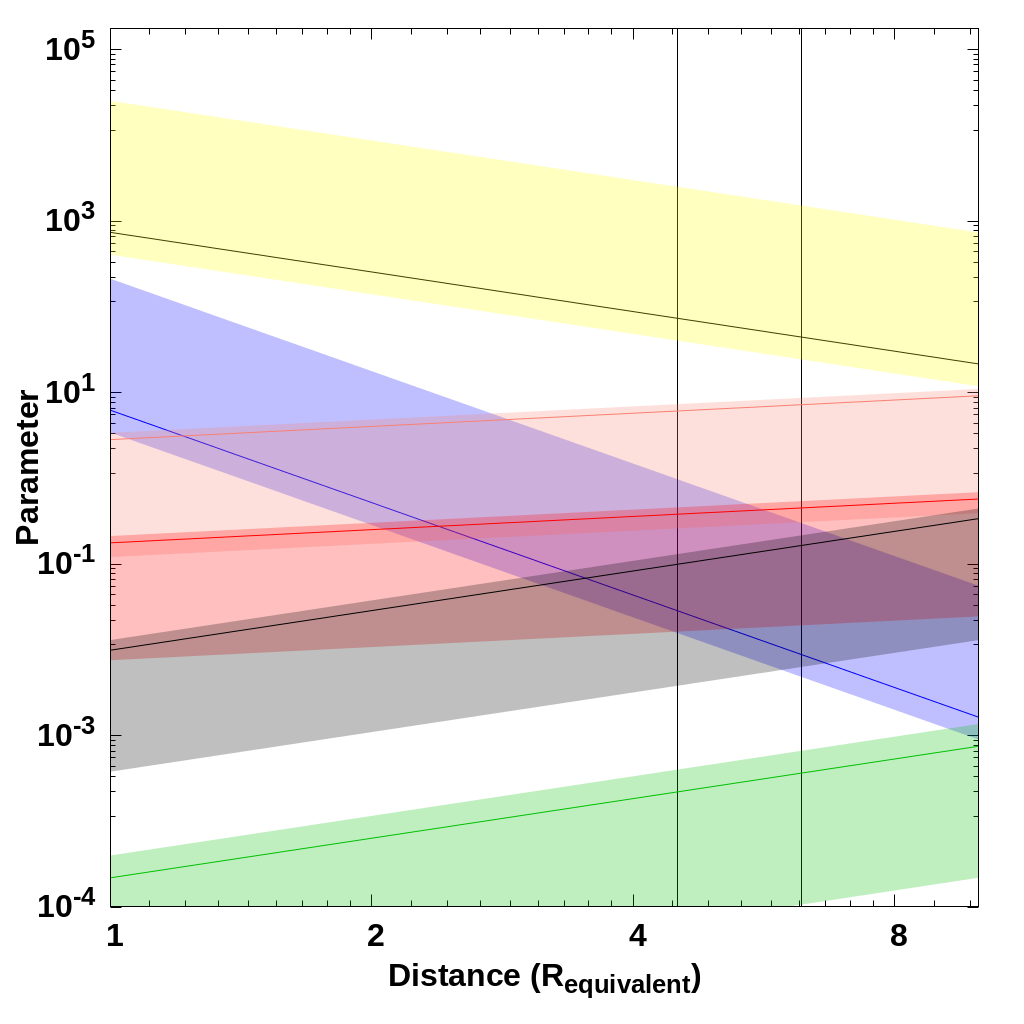}
\end{center}
\caption{\label{fig:param} Variation of the parameters oblateness (blue), Earth and Sun tide (black and green, respectively), the solar radiation pressure for a particle with a radius of 1 cm (pink) and 15 cm (red) and the Apophis gravity (yellow) as a function of the distance from Apophis. The shaded region indicates the variation of these parameters between the pericentre and apocentre. The vertical lines represent the limits of the radial location of the the equilibrium points for the density model of 1.29 g$\cdot$cm$^{-3}$ and 3.5 g$\cdot$cm$^{-3}$. The equivalent radius, $R_{equivalent}$, of Apophis is 170 metres.}
\end{figure}

There have been some studies \citep{aljbaae2020close, aljbaae2021influence, lang2021spacecraft} investigating suitable orbits around Apophis for a spacecraft taking into account the irregular gravitational field of the asteroid and the solar radiation pressure for an area to mass ratio similar to the OSIRES-REX spacecraft. Some stable orbits were found, but during the 2029 close encounter, the majority of the orbits for a spacecraft suffer a collision or ejection.

The small particles around Apophis may be subject to a plethora of forces besides the gravitational potencial of the main body: the oblateness coefficient add an extra gravitational pull and there are the tides by the Sun and the Earth at the close encounter. If the grains are small they can also experience the disturbance due to the solar radiation force.

To estimate which perturbation may be relevant, we computed adimensional parameters that allow us to analyze the relative strength of each force (for a detailed definition of the parameters, see \citet{hamilton1996circumplanetary, moura2020dynamical}). Figure \ref{fig:param} shows the parameter strengths according to the distance from Apophis: the solar tide (green), Earth’s gravity at the closest approach distance (black), the solar radiation force for grains with radius of 1 cm (pink) and 15 cm (red); it is also shown the oblateness effect (blue) and the Apophis gravity (yellow).  For each parameter the lines are calculated for the nominal density (2.2 g$\cdot$cm$^{-3}$) and distance equal to the semimajor axis, while  the shaded region corresponds to the variation of these parameters between the pericentre and apocentre.

In Fig. \ref{fig:param} the vertical lines indicate the limits of the radial location of the equilibrium points for the density model of 1.29 g$\cdot$cm$^{-3}$ and 3.5 g$\cdot$cm$^{-3}$. The Apophis gravity has the major magnitude among the other perturbations, about seven orders of magnitude larger than the Sun tide and five orders larger than the Earth tide at the closest approach distance. The oblateness of Apophis as the Sun tide reaches the same magnitude at a distance of about 9 radii of Apophis. The Earth tide is also lower than the oblateness near the asteroid and they both are comparable in the region of the equilibrium points. Beyond this distance the Earth tide dominates the dynamic over the oblateness. So, the Sun tide will not contribute with major perturbations to the system.

The solar radiation pressure for a particle with a 1 cm of radius gets a larger order of magnitude than the oblateness at about 3 radii of Apophis, what means that this particle will suffer a major influence of the solar radiation pressure. However, for a particle with a 15 cm of radius, the solar radiation pressure is about one order of magnitude smaller, and it is equivalent to the oblateness near the region of the equilibrium points. Thus, a particle of 15 cm is more likely to survive in the nearby environment.

In the current work we are concerned with natural objects, such as dust, boulders, fragments or even larger bodies that might be orbiting around Apophis. Therefore, aiming to identify the size and location of possible stable regions around Apophis and/or around the equilibrium points, we performed sets of numerical simulations of particles around these regions. We numerically integrated a disc with 15,000 massless particles initially for 24 hours (in order to compare to the close encounter with the Earth in 2029 showed in Fig. \ref{fig:orbit_encounter}) and subsequently for 30 years ($\sim$10,000 times the rotation period of the asteroid). 

The particles were distributed at an initial distance of 300 metres from the asteroid centre of mass and different widths for each density model, since the position of the equilibrium points changes as the density changes. For the 1.29 g$\cdot$cm$^{-3}$, 2.2 g$\cdot$cm$^{-3}$ and 3.5 g$\cdot$cm$^{-3}$ densities, the final distances were 1.0 km, 1.15 km, and 1.3 km, respectively. All the particles were assumed to be initially with Keplerian angular velocity in the equatorial plane and circular orbits.

To perform these simulations, we used an N-body integrator package called $N$-$BoM$ \citep{moura2020dynamical, Winter2020}. It considers the gravitational potential of an irregular body as a mass concentration model, MASCONS \citep{Geissler1996}. 
To reproduce the Apophis' gravitational potential field, we calculate the sum of the gravitational potential of all the masses points as \citep{borderes2018}
\begin{equation}
 U (x, y, z)= \sum_{i=1}^{N} \frac{Gm}{r_i},
 \label{eq: mascon}  
\end{equation}
where $N$ is the number of mass points (which is 20,457 in our model), $r_i$ is the distance mascon-particle, and $m$ is the mass of each mascon. As the sum of all mascons is equal to the Apophis' total mass, $M_{Apophis}=Nm$, the mass of each mascon is different according to the density model. 

We considered a particle as ejected if it reaches a distance larger than ten times the distance between the Apophis' centre of mass and the equilibrium point, this is almost equivalent to half of the Hill radius \citep{hamilton1992orbital} of Apophis with respect to the Sun. For the 24 hours simulations, an additional criterion of ejection was considered since some particles may not have sufficient time to exceed the ejection distance due to the short time of the simulation. So we also considered as ejected those particles with positive energy. In this way, a region is called stable where the particles remain at the final integration time, obviously, without collision with the body or be ejected.

\begin{figure*}
\begin{center}
\subfloat[$\rho$ = 1.29 g$\cdot$cm$^{-3}$]{\includegraphics*[trim = 0cm 7cm 7cm 0mm,
width=0.66\columnwidth]{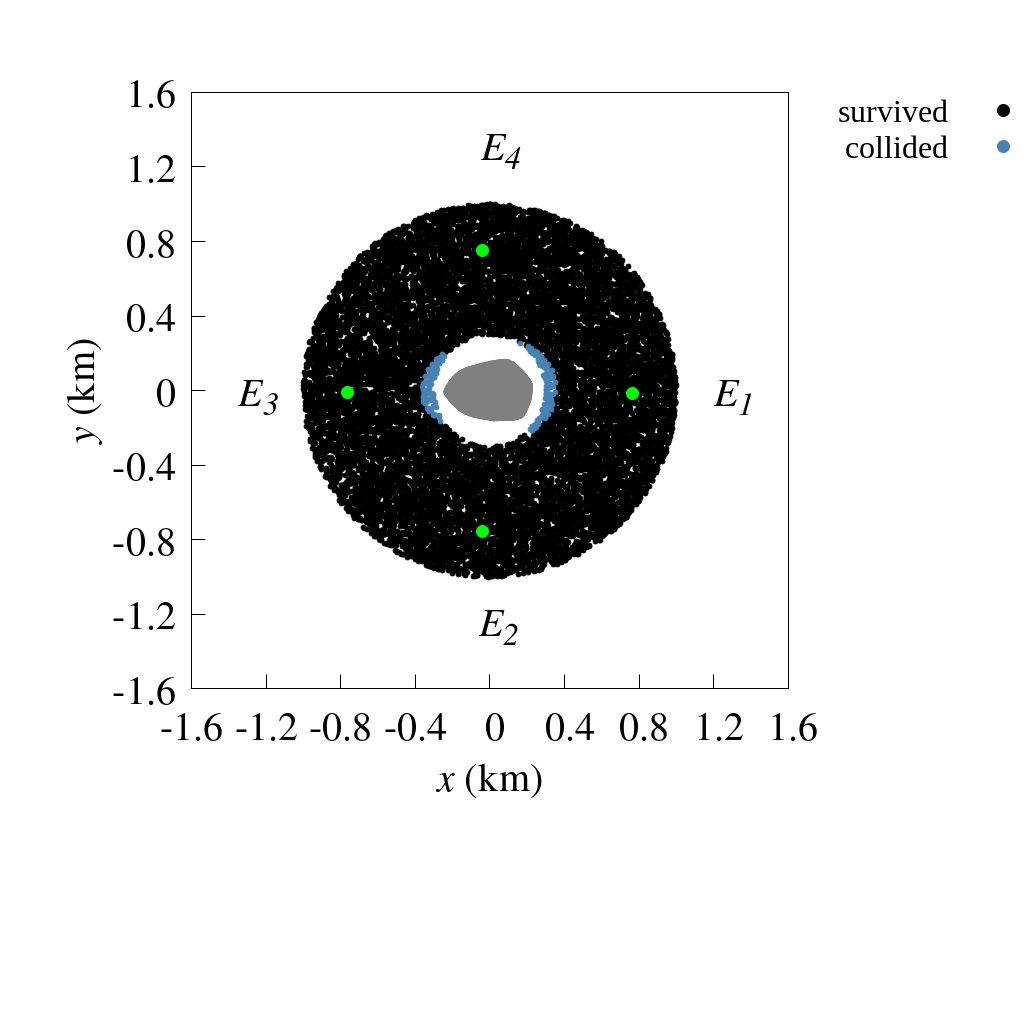}\label{afig:24h_129}}
\subfloat[$\rho$ = 2.2 g$\cdot$cm$^{-3}$]{\includegraphics*[trim = 0mm 7cm 7cm 0mm,
width=0.66\columnwidth]{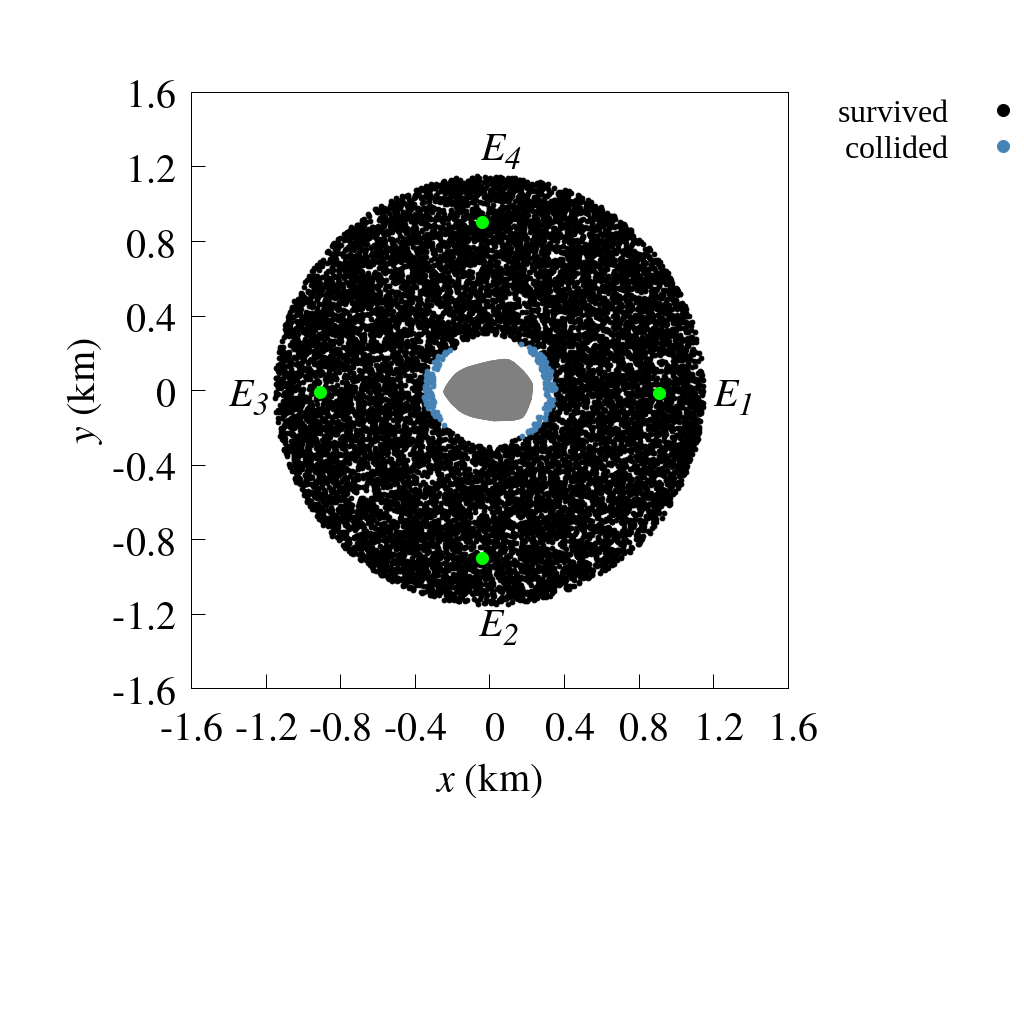}\label{bfig:24h_22}}
\subfloat[$\rho$ = 3.5 g$\cdot$cm$^{-3}$]{\includegraphics*[trim = 0mm 7cm 7cm 0mm, width=0.66\columnwidth]{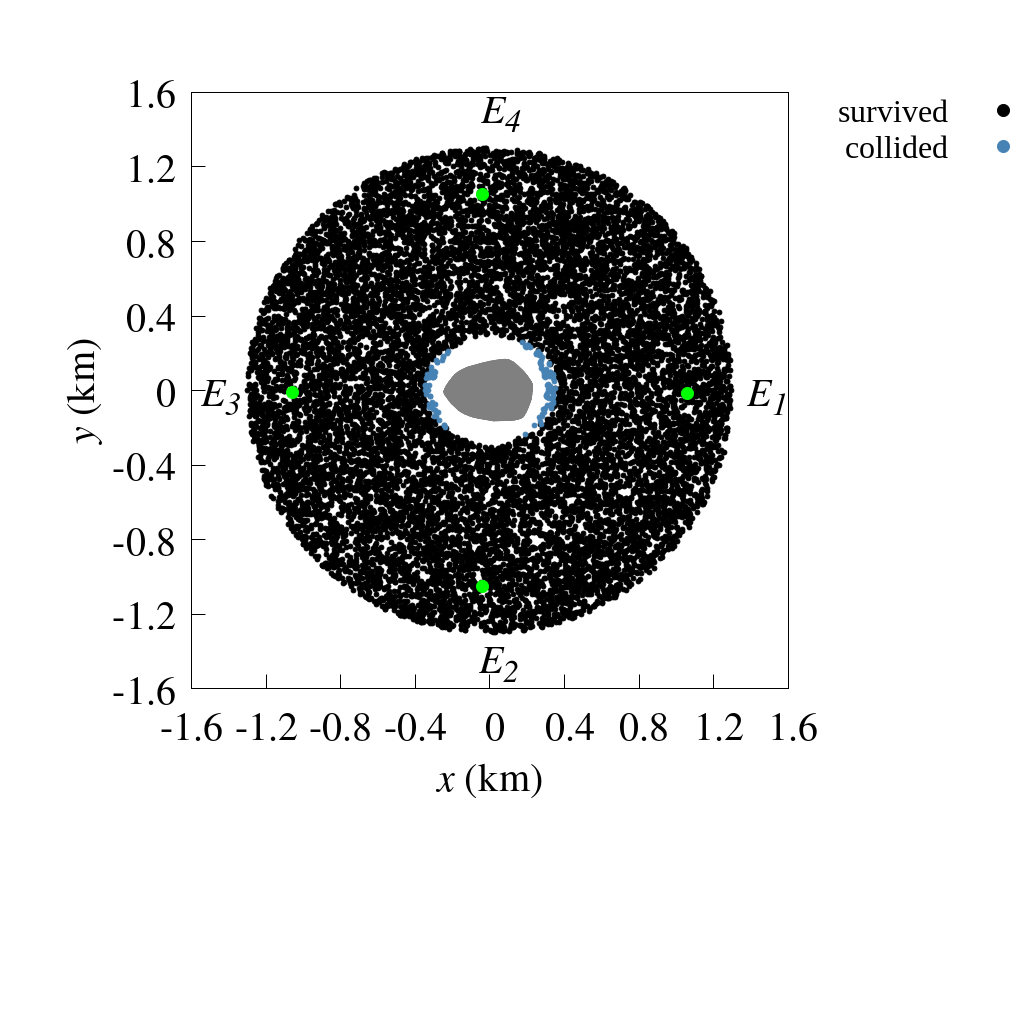}\label{cfig:24h_35}}
\end{center}
\caption{\label{fig:24h} Initial conditions of the 15 thousand particles around the asteroid Apophis in the $xoy$ plane. The green dots represent the equilibrium points. The black dots represent the particles that survived after the 24 hours of integration and the blue dots the particles that collide with the asteroid. The letters (a)-(c) represent, respectively, the densities of 1.29 g$\cdot$cm$^{-3}$, 2.2 g$\cdot$cm$^{-3}$ and 3.5 g$\cdot$cm$^{-3}$.}
\end{figure*}

\begin{figure}
\begin{center}
\subfloat[$\rho$ = 1.29 g$\cdot$cm$^{-3}$]{\includegraphics*[trim = 0mm 6cm 0cm 1cm,
width=1\columnwidth]{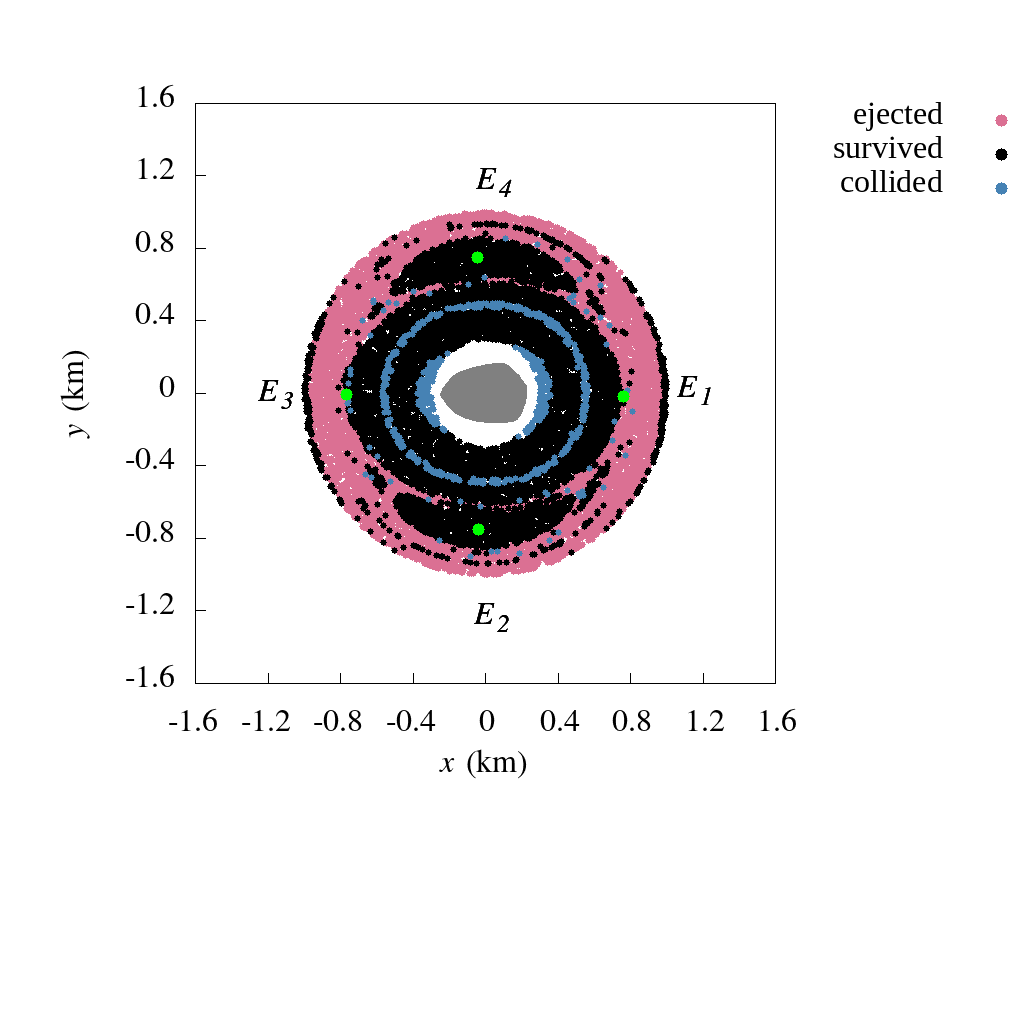}\label{afig:30y_129}}\\
\subfloat[$\rho$ = 3.5 g$\cdot$cm$^{-3}$]{\includegraphics*[trim = 0mm 6.8cm 0cm 1.5cm, width=1\columnwidth]{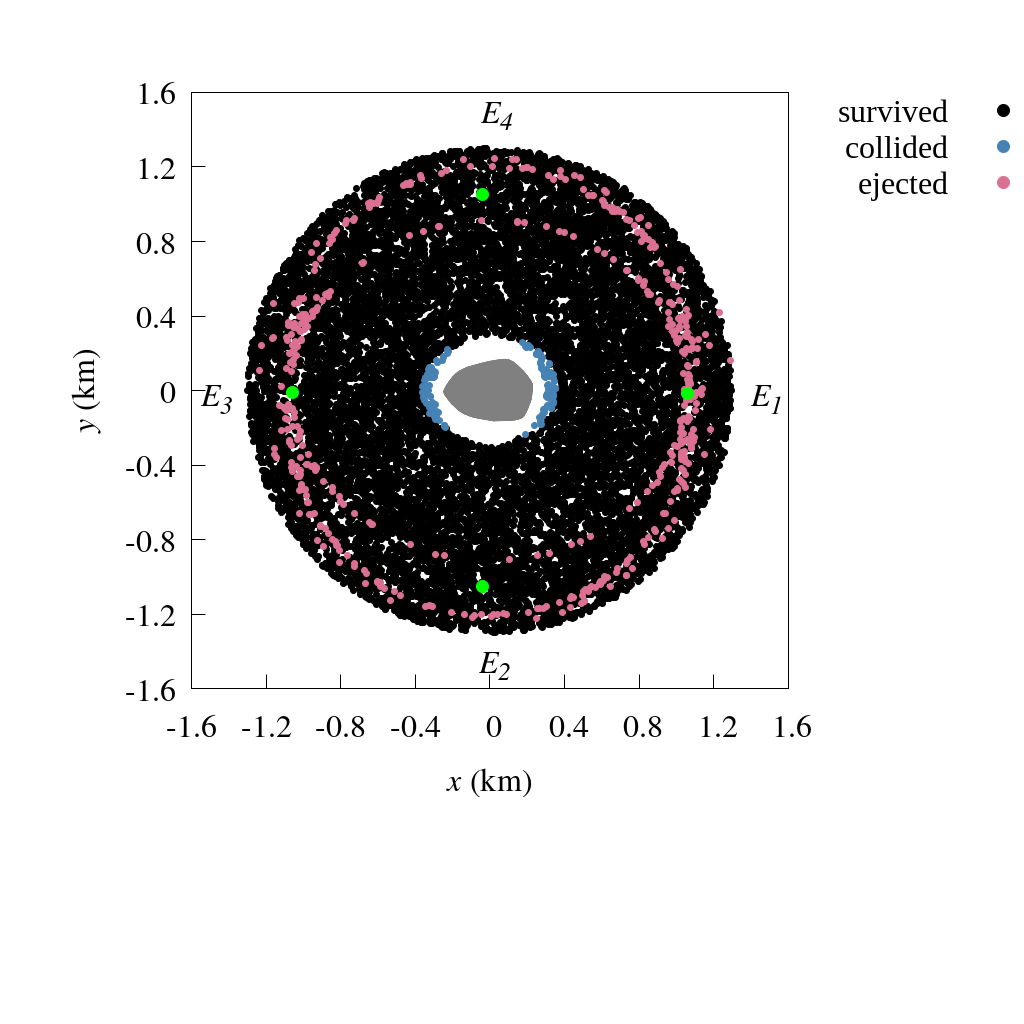}\label{cfig:30y_35}}
\end{center}
\caption{\label{fig:30y} Initial conditions of the 15 thousand particles around the asteroid Apophis in the $xoy$ plane. The green dots represent the equilibrium points. The black dots represent the particles that survived after 30 years, the blue and pink dots are the particles that collide with the asteroid and are ejected from the system, respectively. The letters (a) and (b) represent, respectively, the densities of 1.29 g$\cdot$cm$^{-3}$ and 3.5 g$\cdot$cm$^{-3}$.}
\end{figure}

Figure \ref{fig:24h} shows the initial conditions of the 15,000 particles, colored according to their final outcome at the end of the 24 hours of integration. The blue dots represent the particles that collided with Apophis, while the black dots indicate the particles that have survived the entire simulation. We note that the distribution of the particles that collided with the surface is similar among the three density models, and they are located in two preferential regions closer to the surface of Apophis and near to the equatorial extremities of the body, so the gravitational field of the equatorial extremities is related to the cause of these collisions. 

Most of the particles survived, but we can note a slight difference of the collision percentage among the different density models. For the smaller density, the percentage of particles that collided was about 1.4\%, while for the larger model was only 0.78\%. Although these percentage of collisions, the majority of the disc is stable for the whole time of integration. In section \ref{instability}, we will simulate these same conditions adding the perturbation of the Earth during the trajectory of the 2029 encounter (Fig. \ref{fig:orbit_encounter}) to see its effects in the nearby environment of Apophis.

Extending the simulation for 30 years, we obtain the results shown in Fig. \ref{fig:30y}. The region of collided particles near the asteroid still remains the same for the large density, but an increase of collided particles is noticed for the smaller density. There is a random distribution of collided particles at the proximities of the equilibrium points and a thin ring of collided particles at about 500-600 metres from Apophis. Some of these particles remain in the system for about 10 years and are chaotic, presenting a large orbital radius variation and inclination. 

The ejected particles of the larger density model are distributed in regions involving the equilibrium points, similar to a zero-velocity curve. Note that these particles are located in a limit region where the gravitational and centrifugal forces of the central body exerts influence on them. This means that outside these regions the Apophis influence is sufficiently weak to cause perturbations on the particles, so they survived as we can note in Fig. \ref{cfig:30y_35}. The same behavior is presented in the smaller density model but, since the mass is smaller so the gravitational perturbation exerted on the particles, the ejected regions are larger.

The survived particles are the majority for the larger density model, about 96\% of particles survived for the 30 years simulation, while 3.1\% were ejected and 0.9\% collided with Apophis. Meanwhile, for the smaller density model, 48\% were ejected or collided with Apophis and 52\% survived for 30 years. For both densities, the survived particles are located all over the disc, but there are three preferentially concentrated regions of them (Fig. \ref{fig:30y}). The first one is the region inner the equilibrium points due to the gravitational influence of Apophis. The second one is around both linearly stable equilibrium points $E_2$ and $E_4$. The third region is the region outside the influence of the perturbation of Apophis near the ejected particles, as discussed before. The smaller density model presents some agglomerate of survived particles inside the region of ejected particles, these regions will be studied in detail in future works.

Since the stability region of prograde orbits is about half the Hill radius \citep{domingos2006stable, hunter1967motions}, we expand the radial distribution of the particles around Apophis for the 24 hours simulations. For the smaller, mean, and larger densities, the new radial distances, $r$, were 1.0 $< r <$ 7.5 km, 1.15 $< r <$ 9.0 km, and 1.3 $< r <$ 10.5 km, respectively.

All the 15,000 particles simulated for this new radial distance have survived for the entire 24 hours simulations. The particles are distant from the Apophis, so its irregular gravitational field is low and the system can be interpreted as a two-body problem, thus the region is stable as expected. Later we compare these simulations when the perturbation of the Earth during the 2029 encounter is taken into account.

\subsubsection{Solar Radiation Pressure}
\label{solar}

The previous simulations were performed considering only the gravitational potential of Apophis. However, the solar radiation pressure may cause a significant perturbation on the particles' evolution. To identify the size of the particles that may survive the effects of the solar radiation pressure, we assume particles with different sizes and simulate the same conditions described in section \ref{regions} adding the solar radiation pressure perturbation in the system.

We have carried out numerical simulations with micrometric-sized particles, and the entire ensamble did not survive for the 30 years for particles smaller than 100 $\mu$m. This allowed us to set a lower limit for the particle radius that could survive despite of the solar radiation disturbance. So, we proceed the simulations with the $N$-$BoM$ package for particles with radius larger than 100 $\mu$m. Considering the area to mass ratio ($A$) of the particle and a solar constant in function of the solar luminosity and speed of light ($G^*=1\times10^{17}$ kg$\cdot$m$\cdot$s$^{-2}$), the radiation pressure acceleration is given by \citep{scheeres2002spacecraft}

\begin{equation}
\vec{a}_{srp} = \frac{AG^{*}(1+\eta)}{R^3}\vec{R},
 \label{eq: acel_medium}  
\end{equation}
where $R$ is the module of the position vector Sun-particle ($\vec{R}$) and $\eta$ is the reflectance of the particle, which is assumed to be a unit for a totally reflective material. We do not consider the effects caused by shadowing since they are not significant to produce a considerable change in the surviving scenario, i.e particles that survive the entire simulation without ejection or collision with Apophis.

Since our goal was to determine the order of magnitude of particle's size that could survive in the system despite the perturbation due to the solar radiation force, we simulate a set of particles with radius ($r_p$) in a discrete distribution of 1 cm, 5 cm, 10 cm, and 15 cm. The particles have the same density of Apophis considering each density model (1.29 g$\cdot$cm$^{-3}$, 2.2 g$\cdot$cm$^{-3}$ or 3.5 g$\cdot$cm$^{-3}$).

\begin{figure*}
\begin{center}
\subfloat[$r_p$ = 15 cm and $\rho$ = 1.29 g$\cdot$cm$^{-3}$]{\includegraphics*[trim = 0mm 7cm 7cm 0mm,
width=0.66\columnwidth]{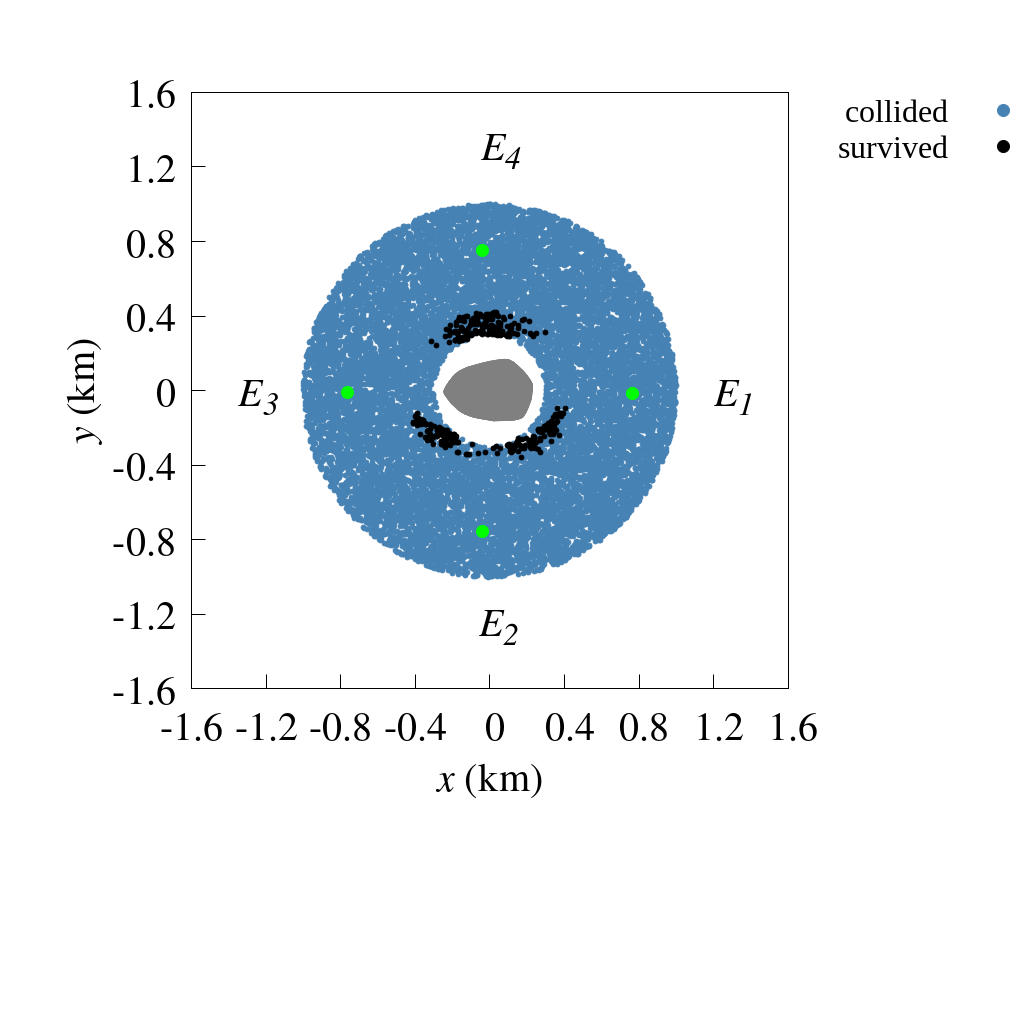}\label{afig:30y_129_srp}}
\subfloat[$r_p$ = 5 cm and $\rho$ = 2.2 g$\cdot$cm$^{-3}$]{\includegraphics*[trim = 0mm 7cm 7cm 0mm,
width=0.66\columnwidth]{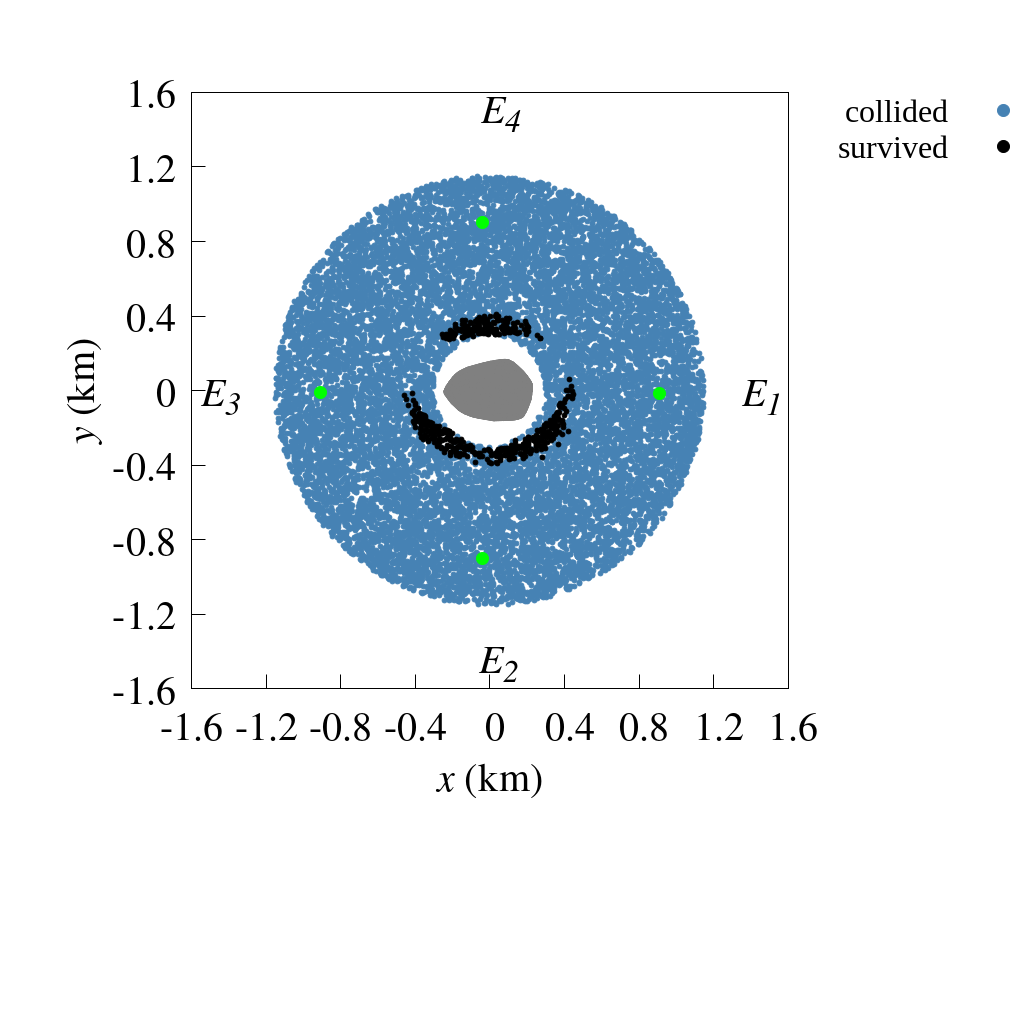}\label{bfig:30y_22_srp}}
\subfloat[$r_p$ = 5 cm and $\rho$ = 3.5 g$\cdot$cm$^{-3}$]{\includegraphics*[trim = 0mm 7cm 7cm 0mm, width=0.66\columnwidth]{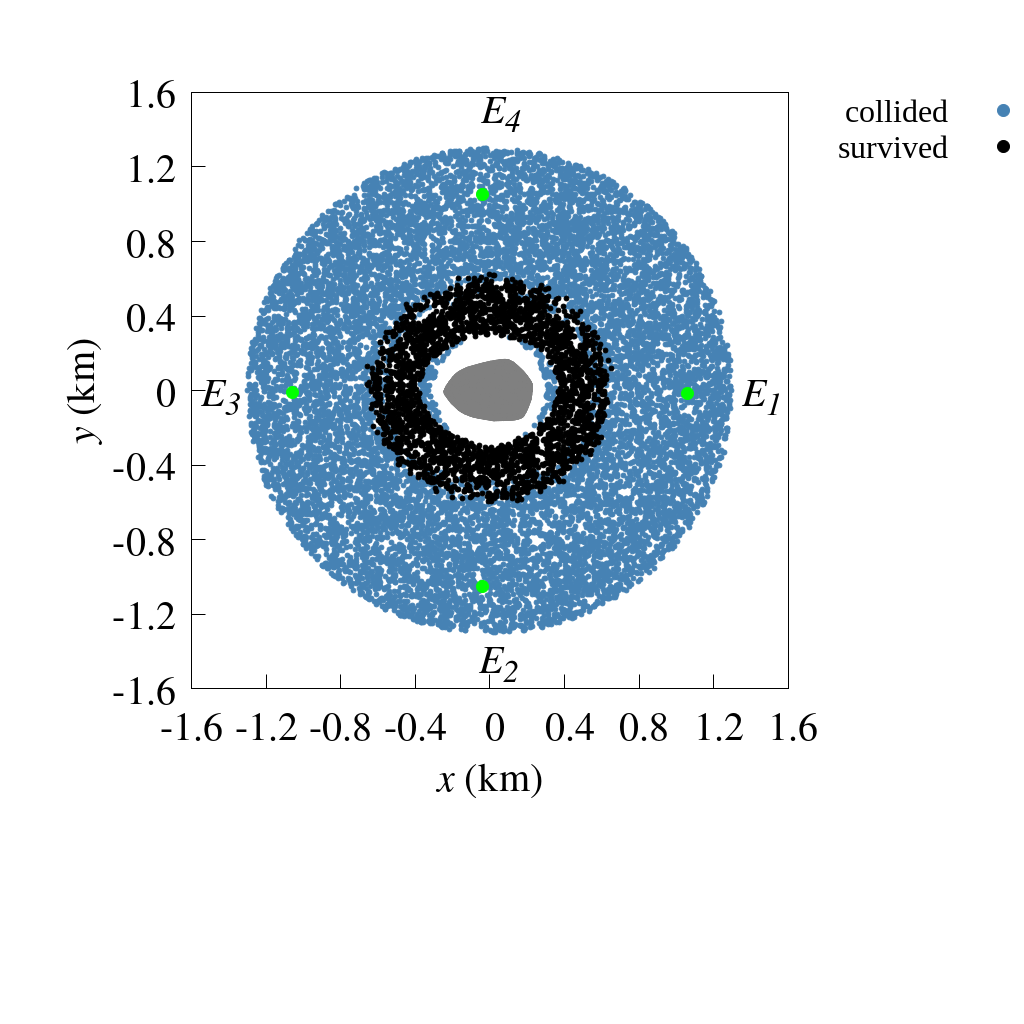}\label{cfig:30y_35_srp}}
\end{center}
\caption{\label{fig:30y_srp} Initial conditions of the 15 thousand particles around the equilibrium points of the asteroid Apophis in the $xoy$ plane considering the perturbation of the solar radiation pressure. The green dots represent the equilibrium points. The black dots represent the particles that survived after the 24 hours of integration and the blue dots the particles that collide with the asteroid. The letters (a)-(c) represent, respectively, the densities of 1.29 g$\cdot$cm$^{-3}$, 2.2 g$\cdot$cm$^{-3}$ and 3.5 g$\cdot$cm$^{-3}$. For the densities of 2.2 g$\cdot$cm$^{-3}$ and 3.5 g$\cdot$cm$^{-3}$ the radius of the particles is 5 cm and for the density of 1.29 g$\cdot$cm$^{-3}$, 15 cm.}
\end{figure*}

For the lower density, Figure \ref{afig:30y_129_srp} shows the initial conditions of the particles where the blue dots represent the collided particles and the black dots the survivors. For the sizes of 1 cm and 5 cm, no particles survived, and for the radius of 10 cm, just one particle survived. Then, we simulated particles with radius of 15 cm, and for this, about 3\% of the 15,000 particles survived 30 years. We notice that the survived particles are concentrated in three small regions near the inner edge of the disc and, consequently, close to Apophis.

For the density models of 2.2 g$\cdot$cm$^{-3}$ and 3.5 g$\cdot$cm$^{-3}$, there are no survivors for 1 cm particles, but for particles with radius of 5 cm it has about 3.5\% of survivors for the mean density model and almost 18\% for the larger one. The small regions with surviving particles for the mean density model are similar to the regions of the smaller model, also presenting three defined regions (Fig. \ref{bfig:30y_22_srp}). However, the larger density model has a single ring of survived particles (Fig. \ref{cfig:30y_35_srp}). Note that we can clearly see the same two preferential regions of collisions close to the  Apophis' surface existent in the 24 hours simulations without the Earth's perturbation (see Fig. \ref{fig:24h}). Thus, just particles with cm-sized survive for 30 years of simulations considering the solar radiation pressure.

\begin{figure*}
\begin{center}
\subfloat[]{\includegraphics*[trim = 0mm 0cm 0cm 0mm,
width=1\columnwidth]{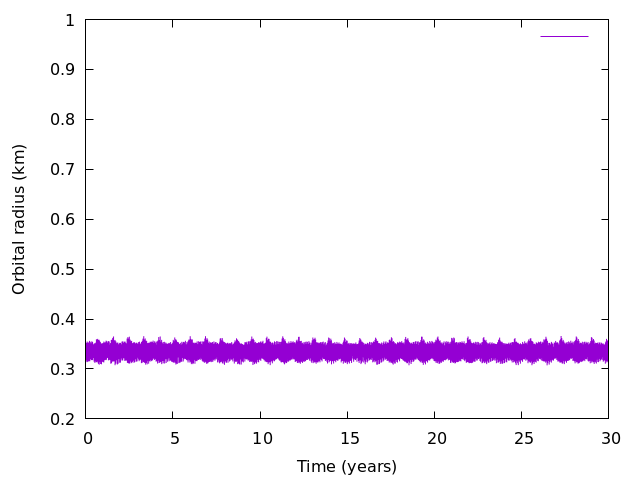}\label{efig:rxt_zxt_sr}}
\subfloat[]{\includegraphics*[trim = 0mm 0cm 0cm 0mm, width=1\columnwidth]{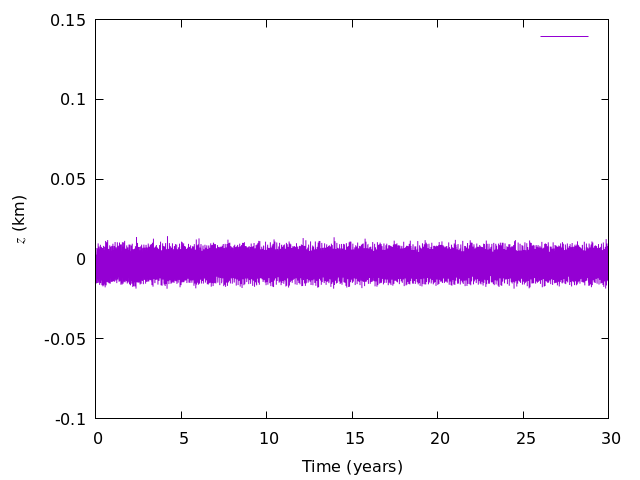}\label{ffig:rxt_zxt_sr}}\\
\subfloat[]{\includegraphics*[trim = 0mm 0cm 0cm 0mm,
width=1\columnwidth]{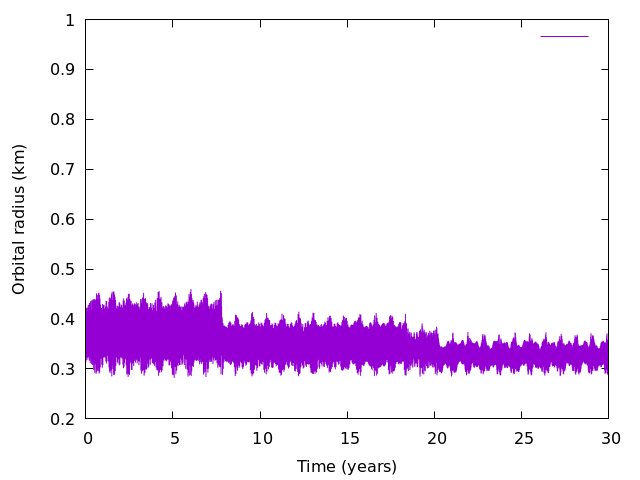}\label{afig:rxt_zxt_sr}}
\subfloat[]{\includegraphics*[trim = 0mm 0cm 0cm 0mm,
width=1\columnwidth]{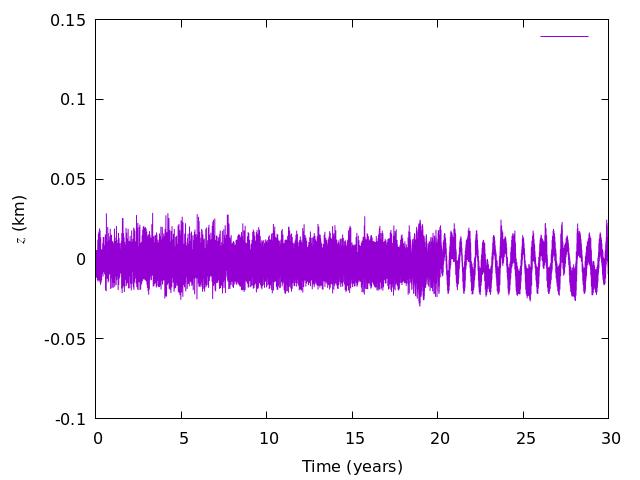}\label{bfig:rxt_zxt_sr}}\\
\subfloat[]{\includegraphics*[trim = 0mm 0cm 0cm 0mm, width=1\columnwidth]{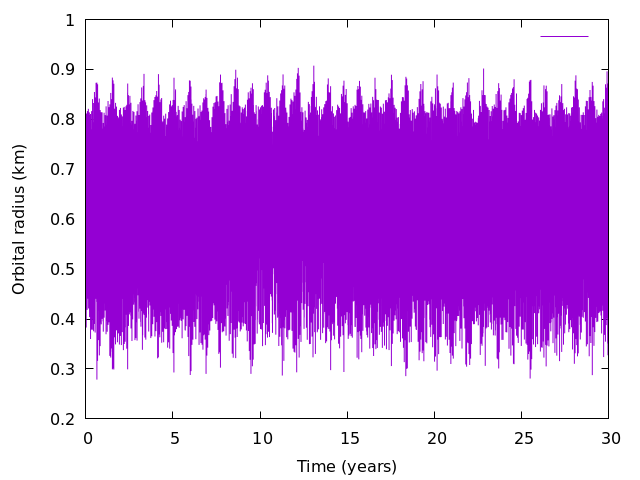}\label{cfig:rxt_zxt_sr}}
\subfloat[]{\includegraphics*[trim = 0mm 0cm 0cm 0mm,
width=1\columnwidth]{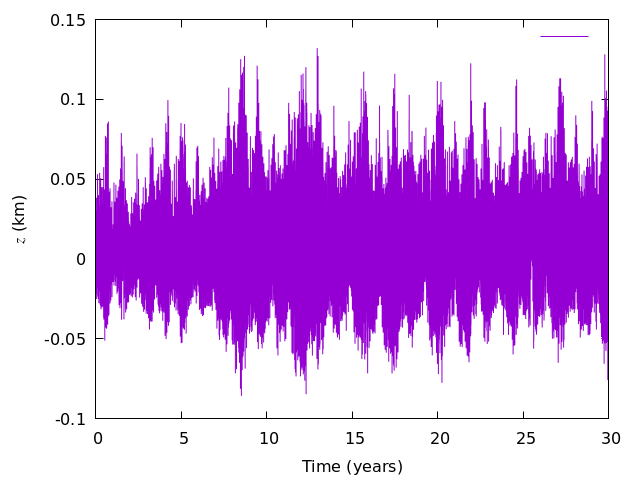}\label{dfig:rxt_zxt_sr}}\\
\end{center}
\caption{\label{fig:rxt_zxt_srp} Three different selected particles among the simulations with solar radiation pressure representing the different orbital radius behavior. The letters (a)-(f) represent the orbital radius and $z$ coordinate for three survived selected particles. The letters (a), (b), (e) and (f) represent the plot for two selected particles with a density of 3.5 g$\cdot$cm$^{-3}$. The letters (c) and (d) represent the plot for a selected particle with a density of 2.2 g$\cdot$cm$^{-3}$.}
\end{figure*}

Analysing the particles that survived for the entire simulation, we note two different predominant behaviours of particles' orbital radius. The first and most common behaviour is a radial oscillation, that could happen either in a large or small scale as exemplified in Figures \ref{cfig:rxt_zxt_sr} and \ref{efig:rxt_zxt_sr}. The small radial oscillation occurs preferentially for the densities models of 1.29 g$\cdot$cm$^{-3}$ and 2.2 g$\cdot$cm$^{-3}$ since the initial conditions of the survived particles are near the Apophis' surface (Figs. \ref{afig:30y_129_srp} and  \ref{bfig:30y_22_srp}). Thus the orbit of these particles could not have a large amplitude, since it would implicate in large eccentricity, a pericentre closer to the surface, and consequently in a collision with Apophis. The small radial oscillation also occurs for the density of 3.5 g$\cdot$cm$^{-3}$, but the frequency of large radial oscillation is larger (Fig. \ref{cfig:30y_35_srp}). Since there is a large region of survived particles and most of them have initial conditions more distant from the body, the orbit of these particles can obtain larger amplitudes and eccentricities without collision with Apophis.

The second behavior is an abrupt radial variation and occurs for the three densities models. The irregular gravitational potential of the body could produce a variation of eccentricity. When the eccentricity increases the particle collides with Apophis and when it decreases, it produces the decays shown in Figure \ref{afig:rxt_zxt_sr}. Despite different behavior, the orbital radius of all the particles that survived are restricted to the minimum value of 300 metres, approximately, which is expected since an inferior value could result in a collision with the body's surface. The $z$ coordinate also has a variation (Figs. \ref{efig:rxt_zxt_sr},  \ref{ffig:rxt_zxt_sr} and \ref{dfig:rxt_zxt_sr}), implying the change in the orbits' inclination. The amplitude of the orbital inclination is related to the amplitude of the orbital variation, thus orbits with larger eccentricity allow larger inclinations.

\section{Instability Due to the Earth's Encounter}
\label{instability}

To understand how the Earth will affect the environment around Apophis in the 2029 encounter, we simulate the initial condition described in section \ref{regions} for the disc encompassing the equilibrium points and the expanded disc for 24 hours adding the perturbation of the Earth and using the trajectory shown in Fig. \ref{fig:orbit_encounter}. Those particles we simulate are at least cm-sized, otherwise they are removed by the solar radiation pressure. 

To evaluate the largest possible perturbation of the Earth, we positioned Apophis in such a way that its orbital plane coincides with its equatorial plane of the Earth. Then, we interpolate this trajectory in the $N$-$BoM$ integrator as an additional external force and simulated the 30 thousand particles around Apophis for 24 hours.

\begin{figure*}
\begin{center}
\subfloat[]{\includegraphics*[trim = 0mm 7cm 6.5cm 0mm,
width=0.66\columnwidth]{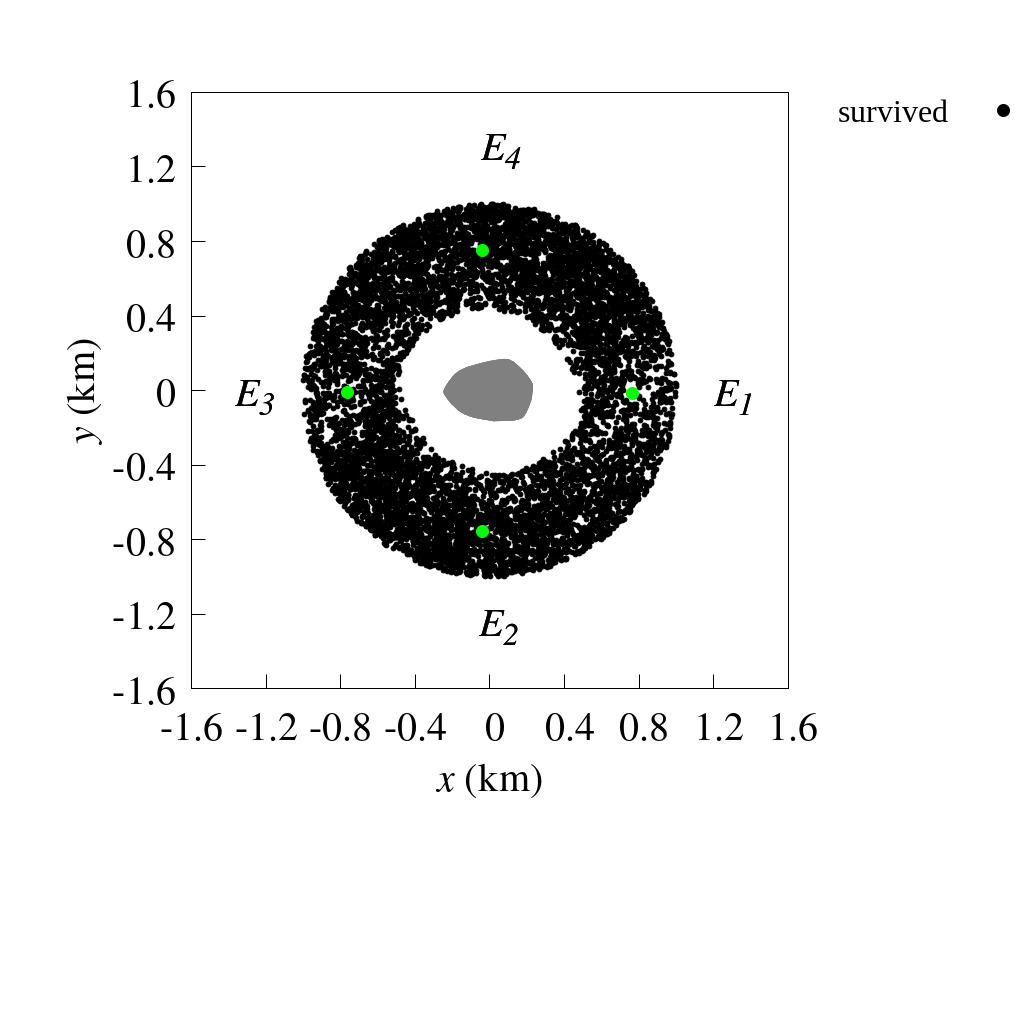}\label{afig:24hT_129}}
\subfloat[]{\includegraphics*[trim = 0mm 7cm 6.5cm 0mm,
width=0.66\columnwidth]{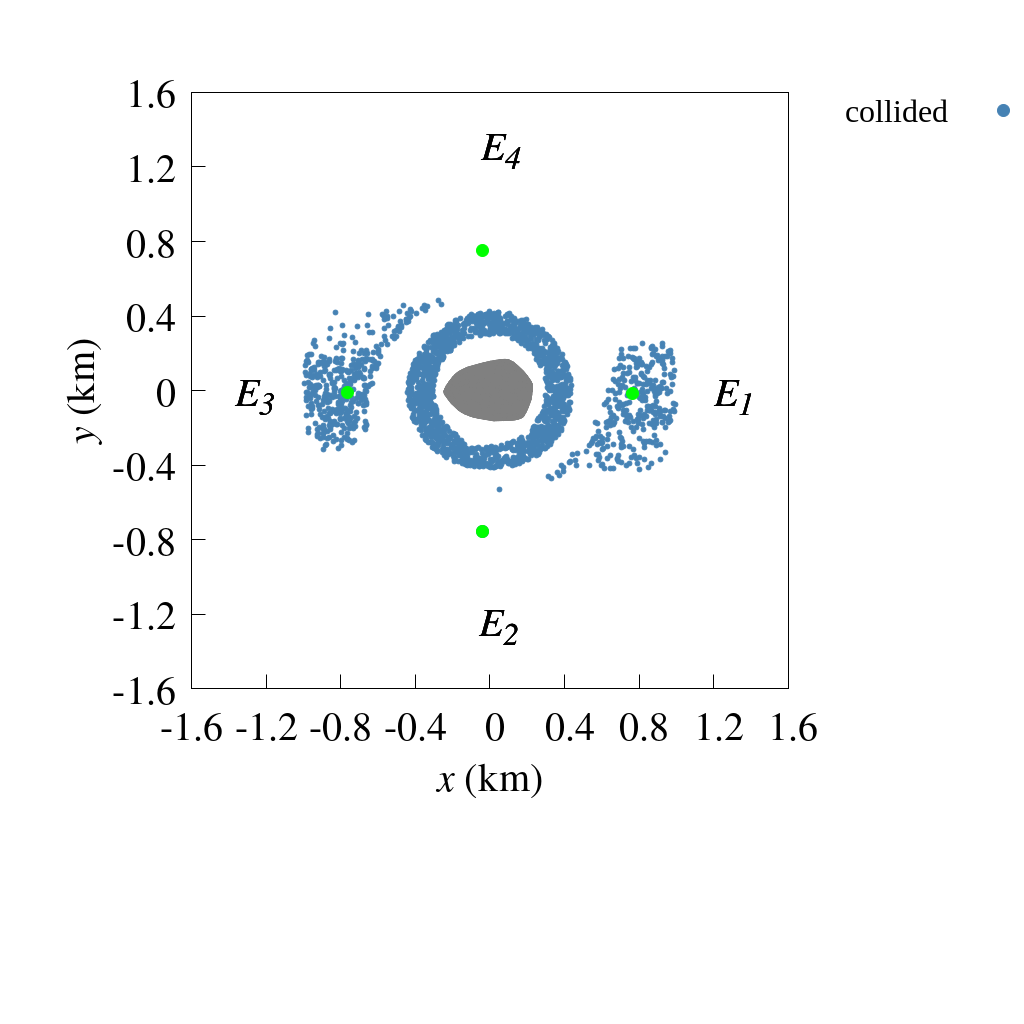}\label{bfig:24hT_22}}
\subfloat[]{\includegraphics*[trim = 0mm 7cm 6.5cm 0mm, width=0.66\columnwidth]{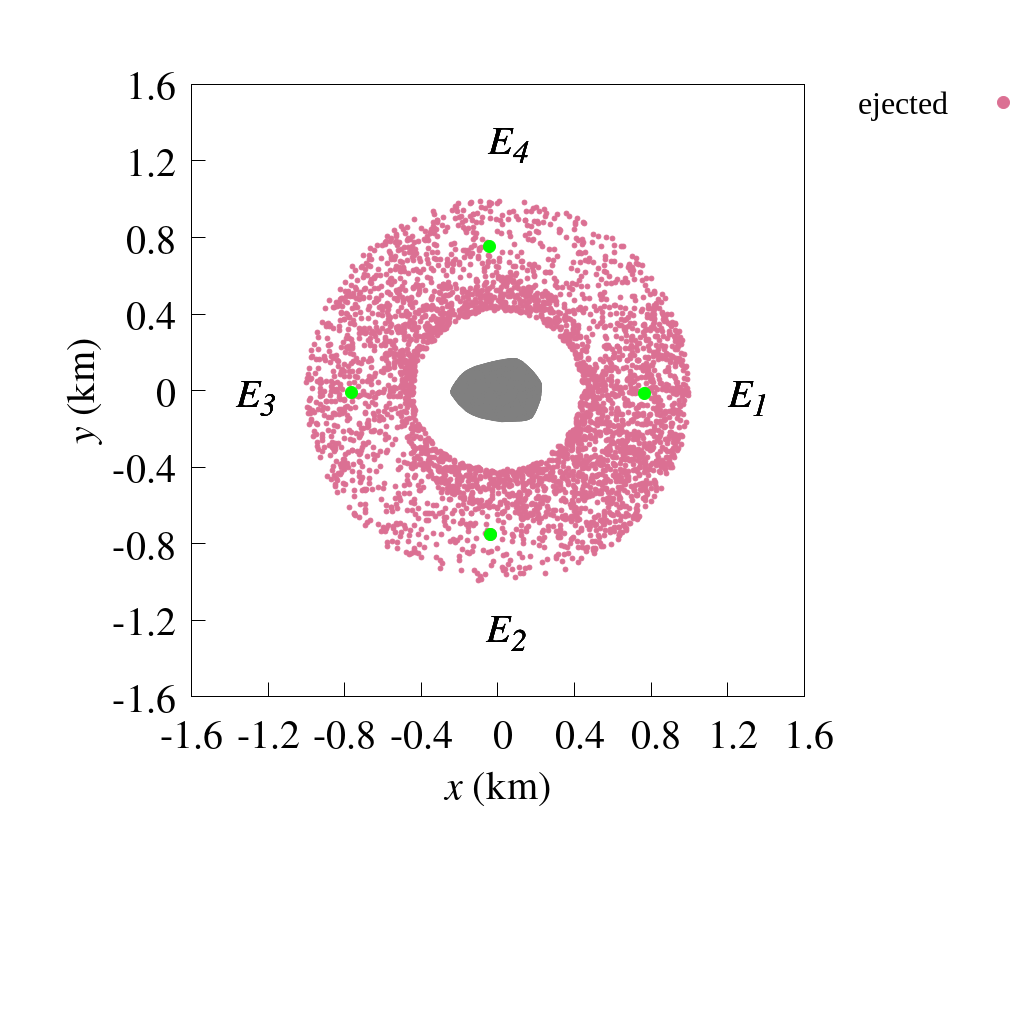}\label{cfig:24hT_35}}\\
\subfloat[]{\includegraphics*[trim = 0mm 7cm 6.5cm 0mm,
width=0.66\columnwidth]{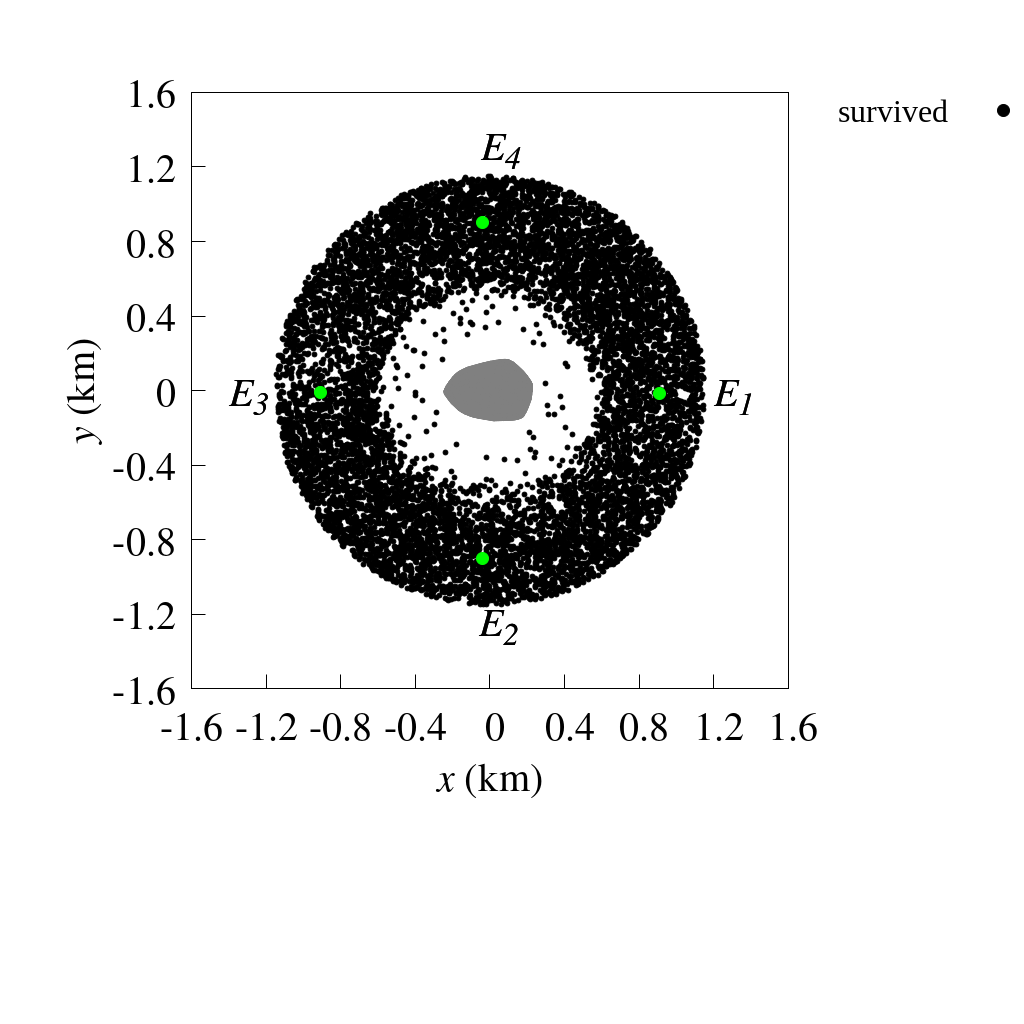}\label{dfig:24hT_129}}
\subfloat[]{\includegraphics*[trim = 0mm 7cm 6.5cm 0mm,
width=0.66\columnwidth]{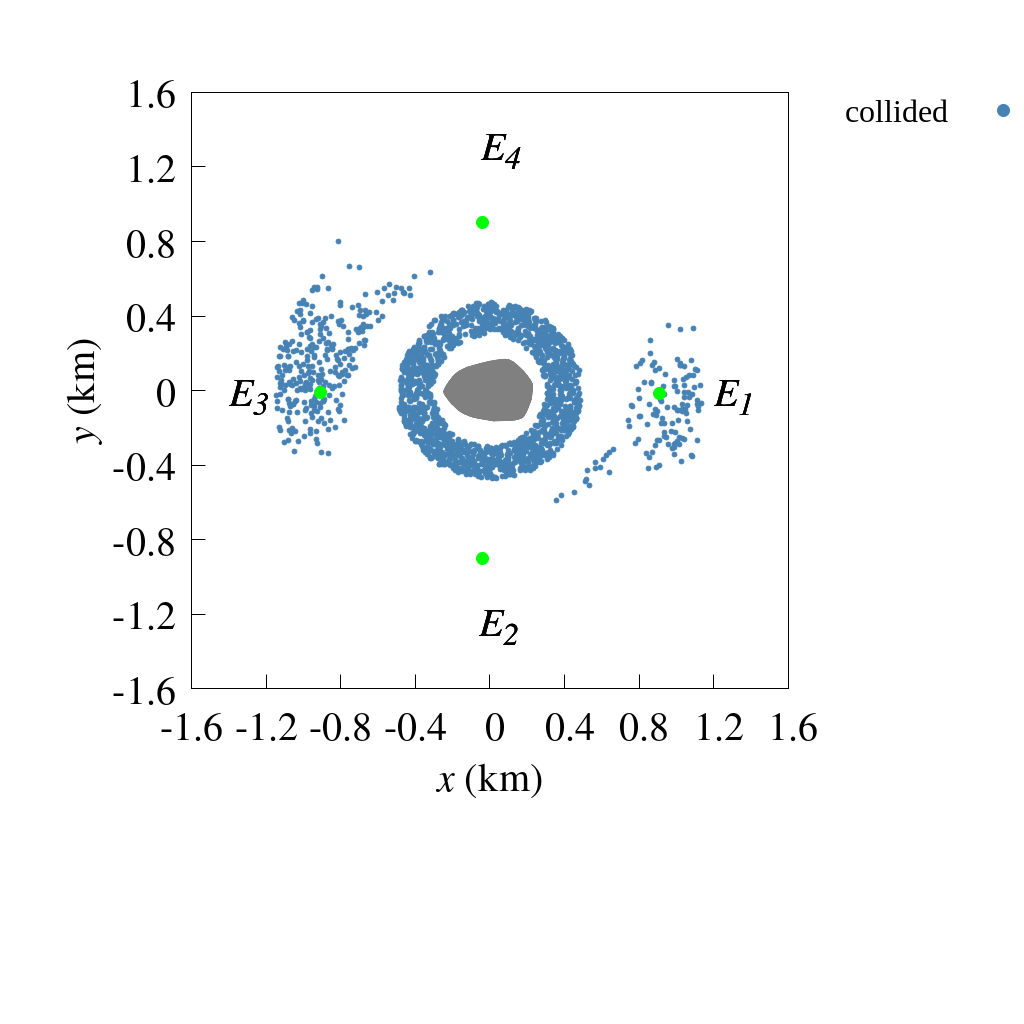}\label{efig:24hT_22}}
\subfloat[]{\includegraphics*[trim = 0mm 7cm 6.5cm 0mm, width=0.66\columnwidth]{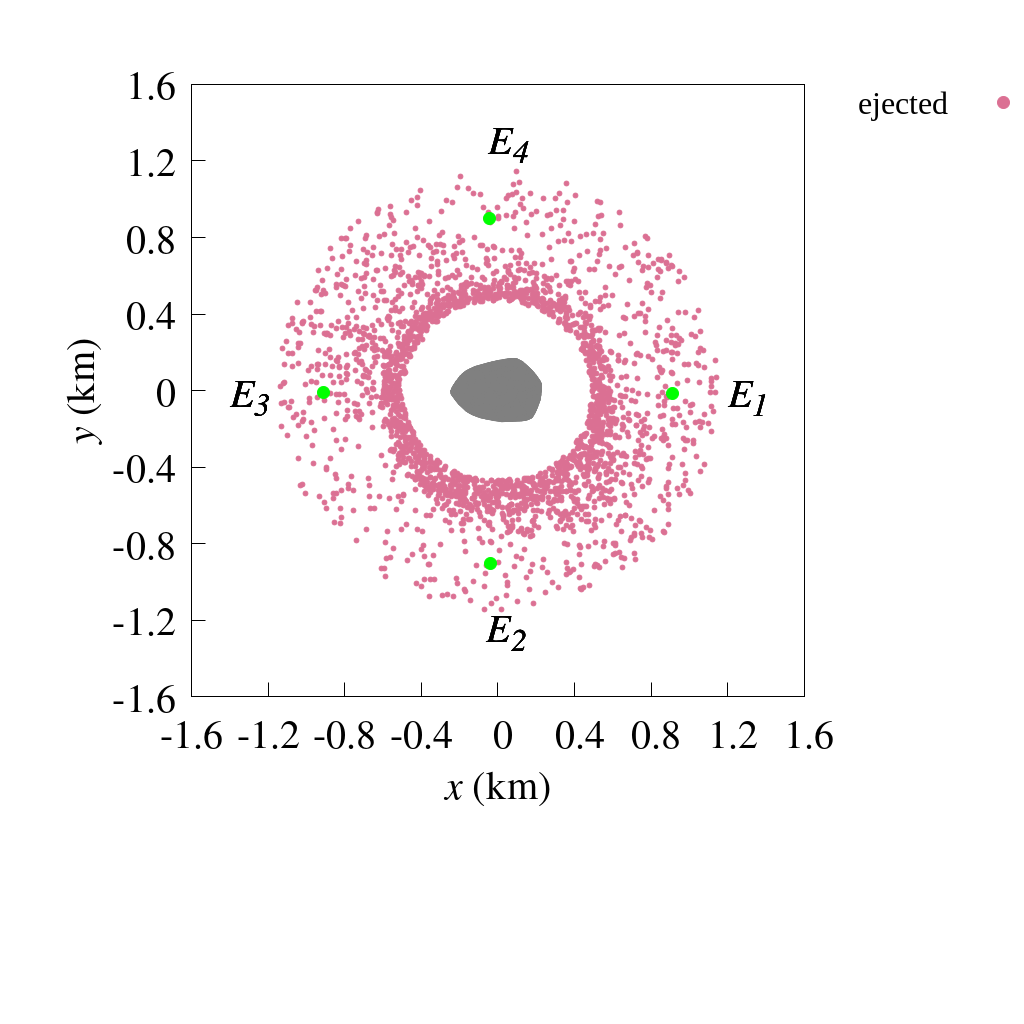}\label{ffig:24hT_35}}\\
\subfloat[]{\includegraphics*[trim = 0mm 7cm 6.5cm 0mm,
width=0.66\columnwidth]{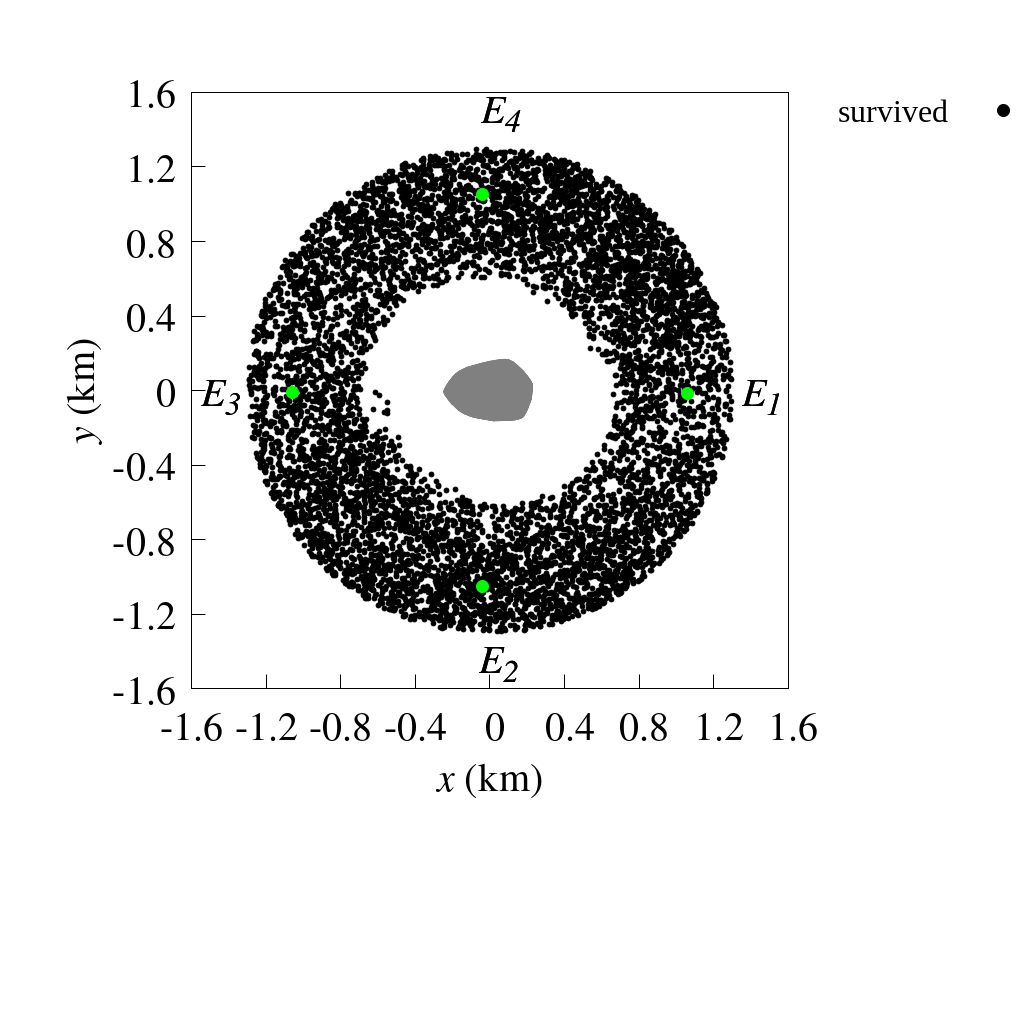}\label{gfig:24hT_129}}
\subfloat[]{\includegraphics*[trim = 0mm 7cm 6.5cm 0mm,
width=0.66\columnwidth]{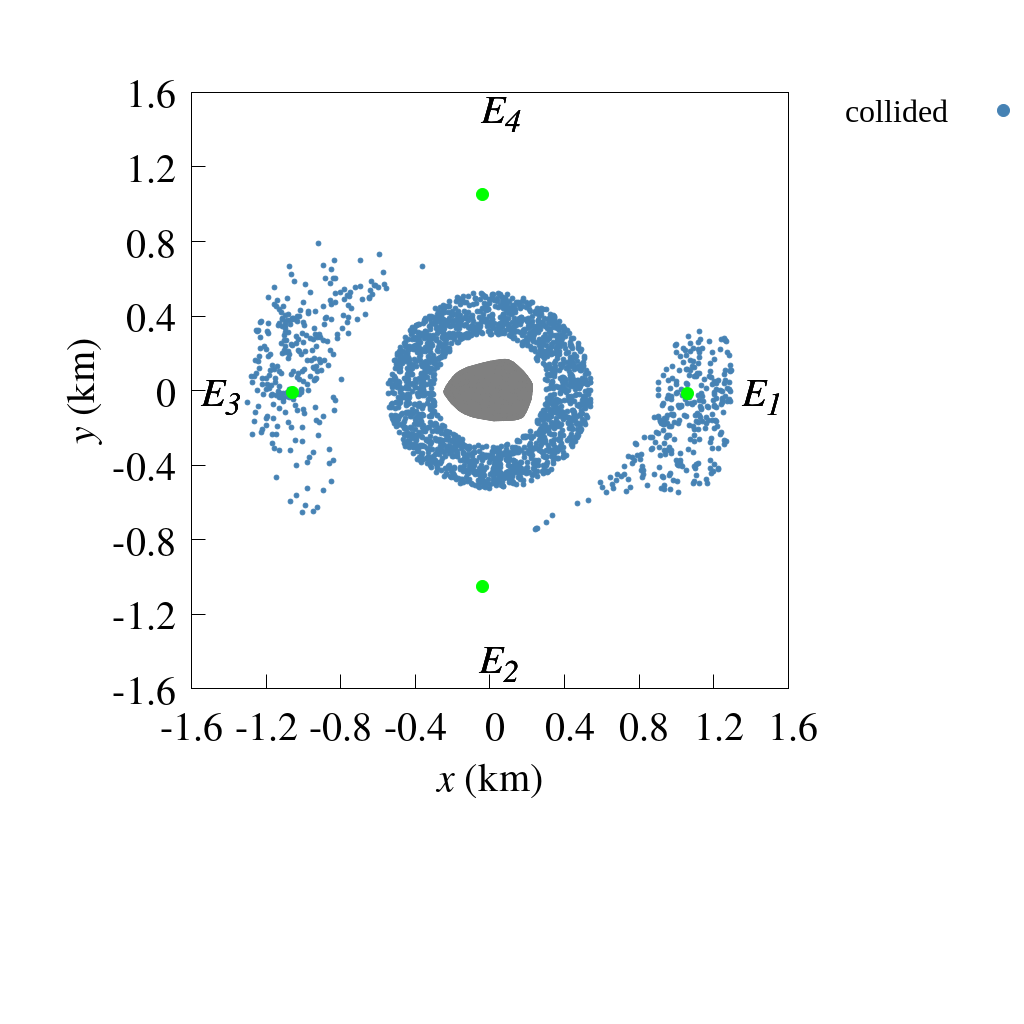}\label{ffig:24hT_22}}
\subfloat[]{\includegraphics*[trim = 0mm 7cm 6.5cm 0mm, width=0.66\columnwidth]{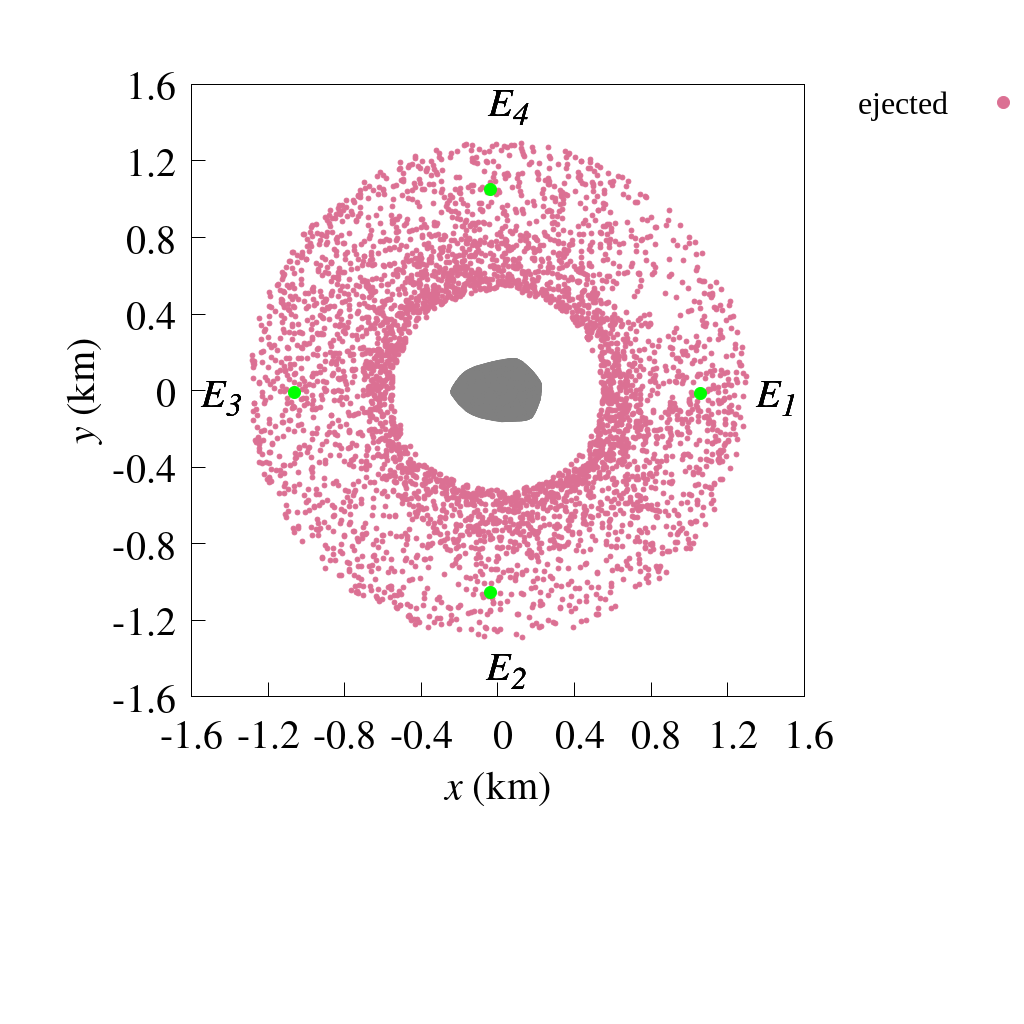}\label{hfig:24hT_35}}
\end{center}
\caption{\label{fig:24hT} Initial conditions of the 15 thousand particles around the equilibrium points of the asteroid Apophis in the 2029 approach in the $xoy$ plane. The green dots represent the equilibrium points. The black dots represent the particles that survived after the 24 hours of integration and the blue and pink dots the particles that collide and ejected, respectively, with the asteroid. The letters (a)-(c), (d)-(f) and (g)-(i) represent, respectively, the densities of 1.29 g$\cdot$cm$^{-3}$, 2.2 g$\cdot$cm$^{-3}$ and 3.5 g$\cdot$cm$^{-3}$.}
\end{figure*}

The initial positions of the 15,000 particles around the equilibrium points in the encounter are shown in Fig. \ref{fig:24hT} for the three densities models. The larger density model has about 56\% of surviving particles, while 16\% and 28\% collided and ejected, respectively. On the other hand, the smaller density model has about 59\% of surviving particles, while 15\% collided and 26\% were ejected by the Earth perturbation.

Without the Earth perturbation, almost 99\% of the particles survive for 24 hours (section \ref{regions}). So, the encounter causes a large change in the Apophis nearby environment as the number of surviving particles decreases approximately 40\% and 44\% for the smaller and larger model of density, respectively. The number of collisions due to the approach increase about 19 times for the smaller model of density and 11 times for the larger model. The ejection of the particles did not occur for the previous simulation, however, considering the 2029 approach 26-28\% of the particles were ejected.

\begin{figure}
\begin{center}
\subfloat[]{\includegraphics*[trim = 0mm 0cm 0cm 0mm,
width=1\columnwidth]{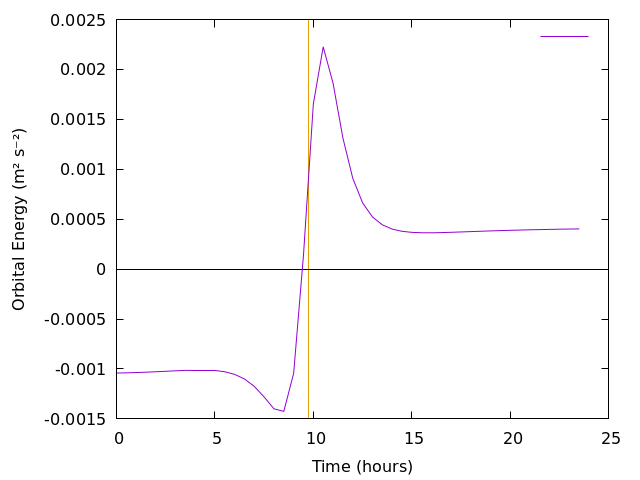}\label{afig:24h_energy4}}\\
\subfloat[]{\includegraphics*[trim = 0mm 0cm 0cm 0mm,
width=1\columnwidth]{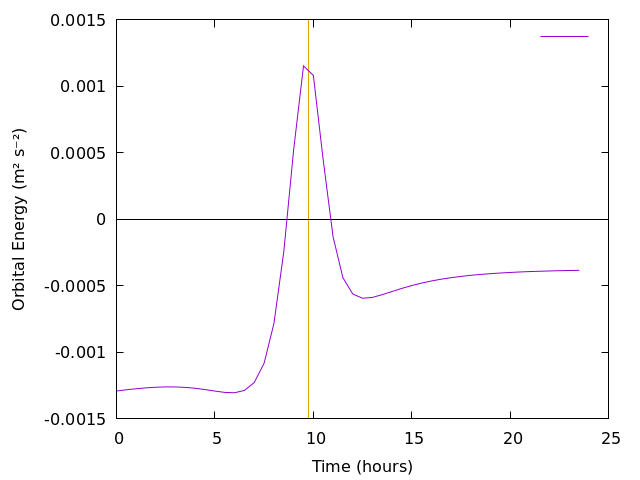}\label{bfig:24h_energy20}}
\end{center}
\caption{\label{fig:24h_energy} Two-body orbital energy of two different particles in the 24 hours simulation considering the 2029 encounter with the Earth. The orange line denotes the time that the closest approach occurs and the black line represents the line of zero energy. The letters (a) and (b) represent the particle that was ejected due to the Earth perturbation and the surviving one, respectively.}
\end{figure}

Some of the ejections occur by the energy criterion, as the Apophis approaches the Earth the orbital energy of the particles increase, and some of them remain positive for the rest of the simulation. Figure \ref{fig:24h_energy} shows the energy of two different particles, a particle that survived and an ejected one. The orange line represents the moment that occurs the closest distance between Apophis and the Earth and the black line represents the line of zero energy. The energy of the particle that was ejected is shown in Fig. \ref{afig:24h_energy4}, observe that near the closest encounter with the Earth the orbital energy increases to approximately 0.0025 m$^2\cdot$s$^{-2}$ and after the encounter, it decreases to 0.0005 m$^2\cdot$s$^{-2}$ until the end of the simulation. Similar behaviour occurs for the particle that survived the 2029 approach (Fig. \ref{bfig:24h_energy20}), but after the encounter, the particle energy decreases to $-$0.0005 m$^2\cdot$s$^{-2}$.

\begin{figure*}
\begin{center}
\subfloat[]{\includegraphics*[trim = 0mm 7cm 6.5cm 0mm,
width=0.66\columnwidth]{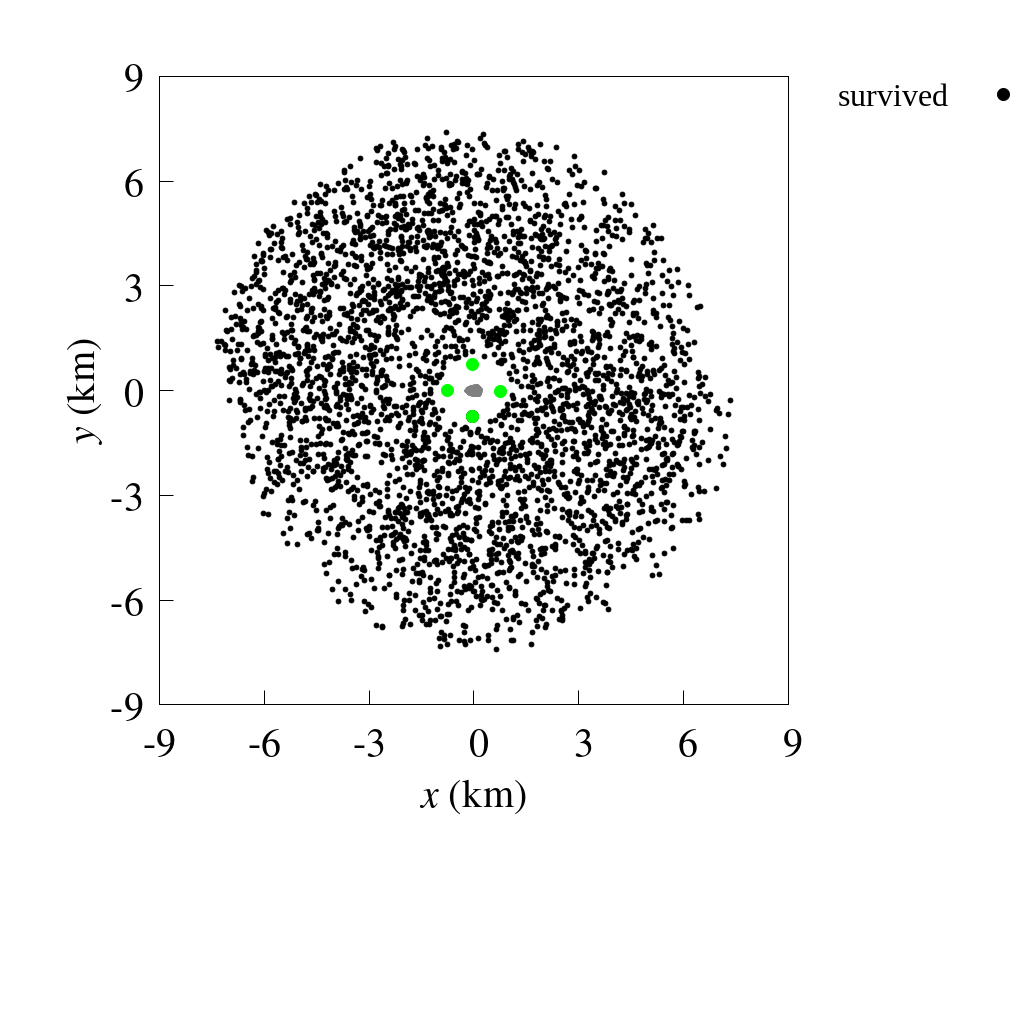}\label{afig:24hT_129L}}
\subfloat[]{\includegraphics*[trim = 0mm 7cm 6.5cm 0mm,
width=0.66\columnwidth]{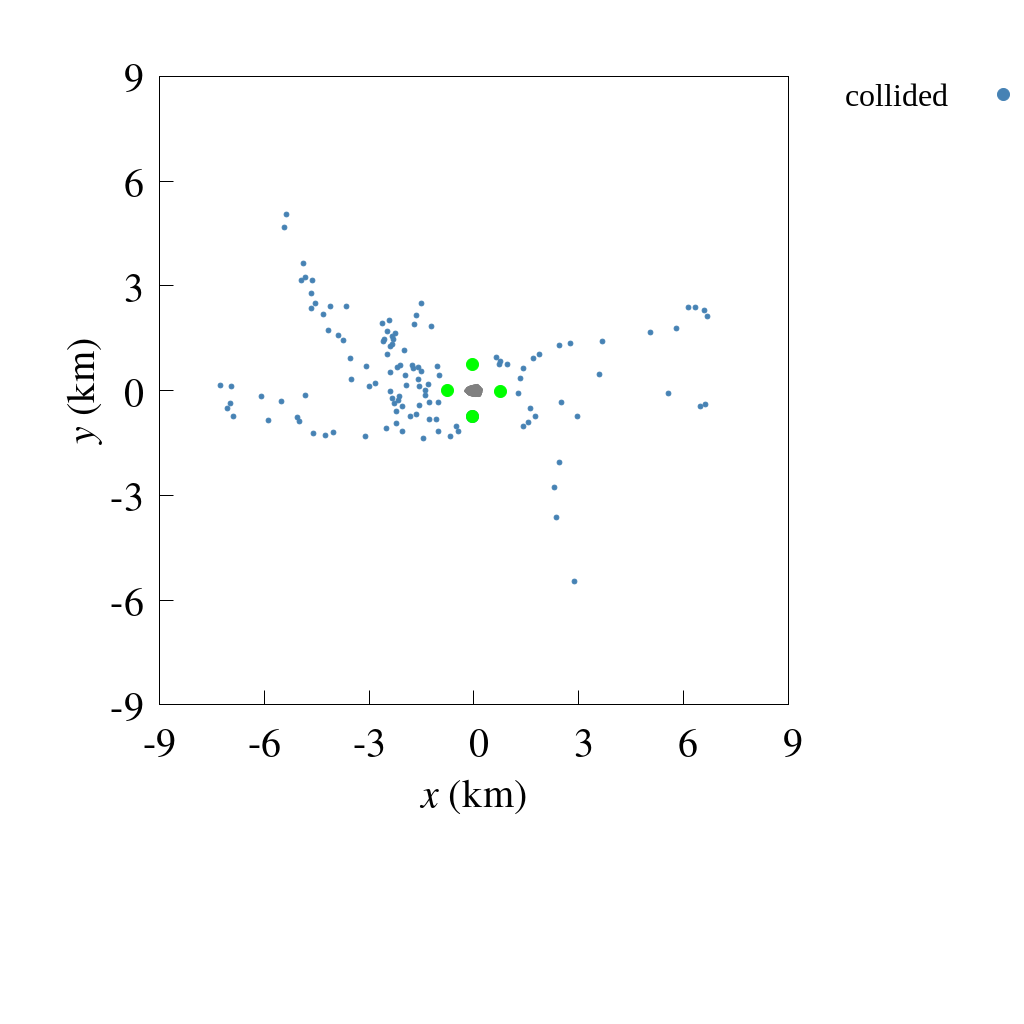}\label{bfig:24hT_22L}}
\subfloat[]{\includegraphics*[trim = 0mm 7cm 6.5cm 0mm, width=0.66\columnwidth]{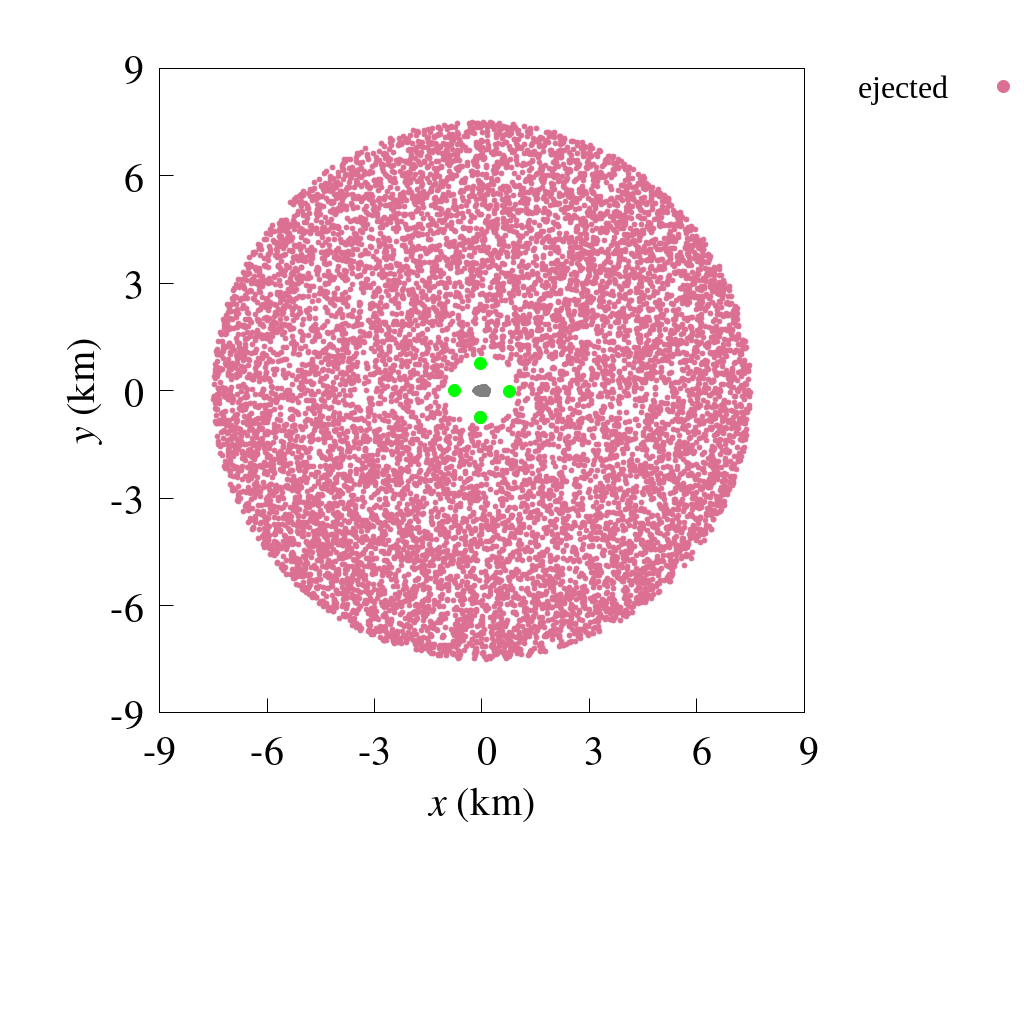}\label{cfig:24hT_35L}}\\
\subfloat[]{\includegraphics*[trim = 0mm 7cm 6.5cm 0mm,
width=0.66\columnwidth]{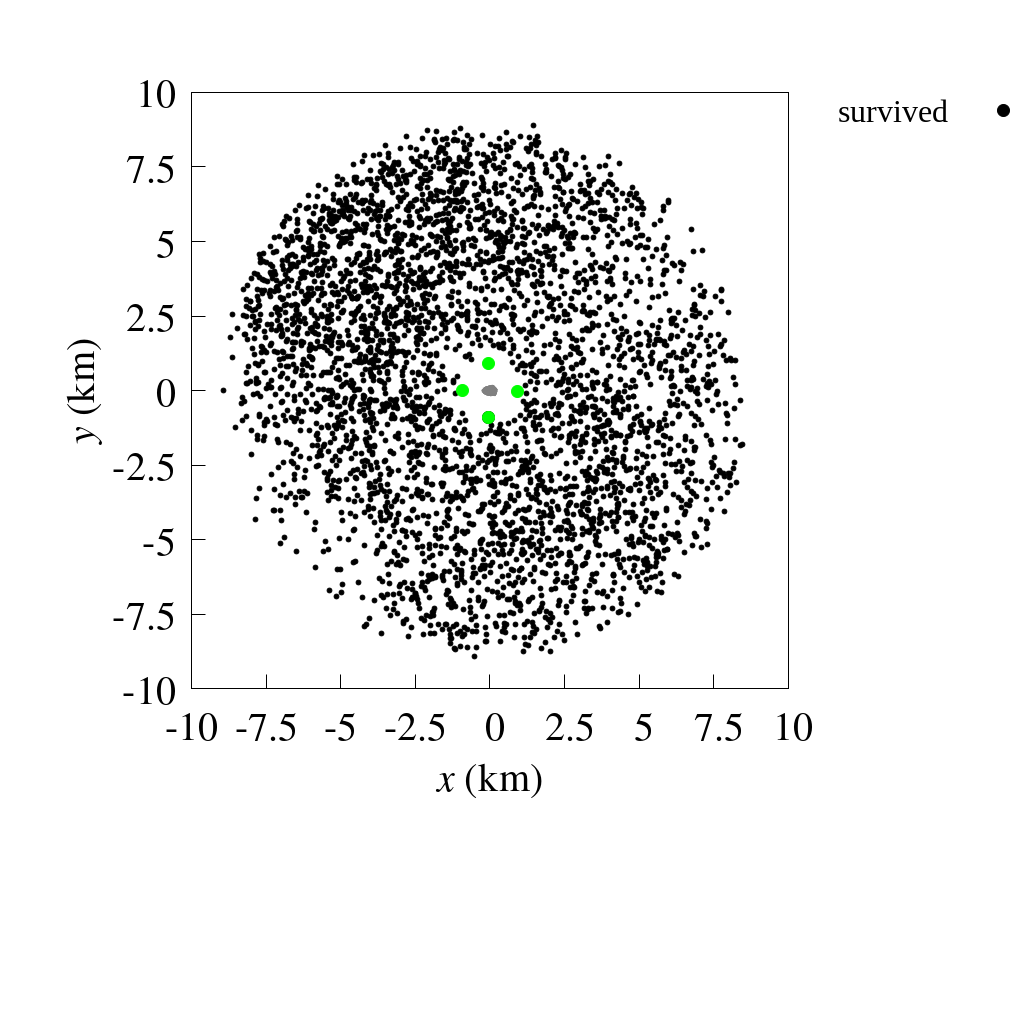}\label{dfig:24hT_129L}}
\subfloat[]{\includegraphics*[trim = 0mm 7cm 6.5cm 0mm,
width=0.66\columnwidth]{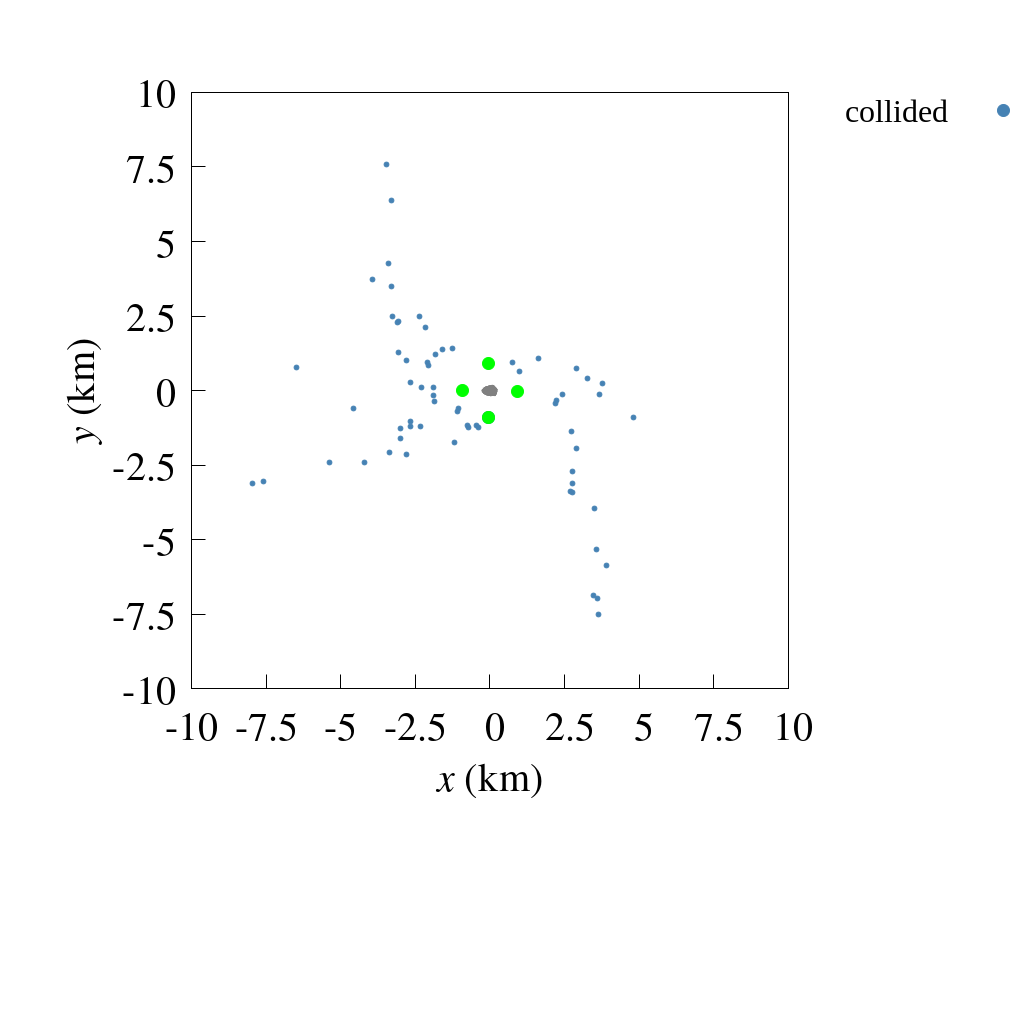}\label{efig:24hT_22L}}
\subfloat[]{\includegraphics*[trim = 0mm 7cm 6.5cm 0mm, width=0.66\columnwidth]{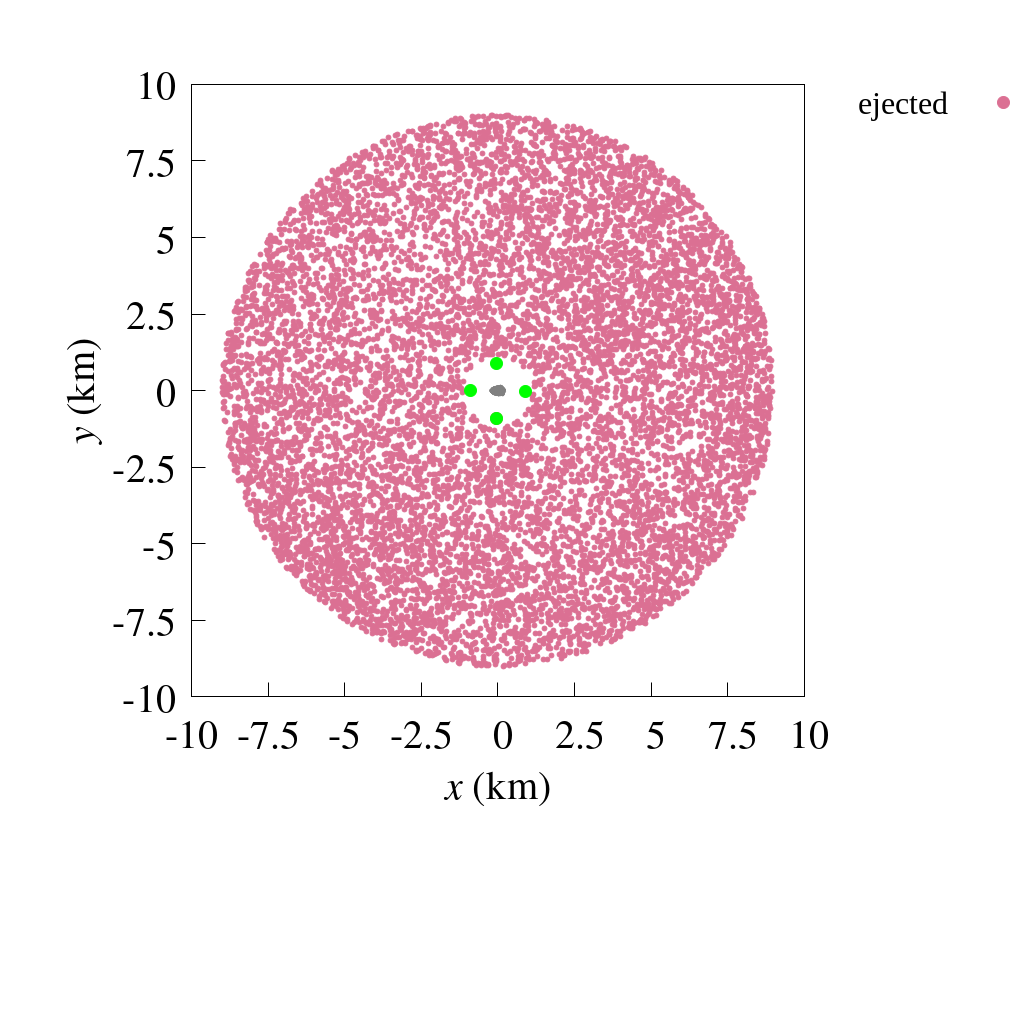}\label{ffig:24hT_35L}}\\
\subfloat[]{\includegraphics*[trim = 0mm 7cm 6.5cm 0mm,
width=0.66\columnwidth]{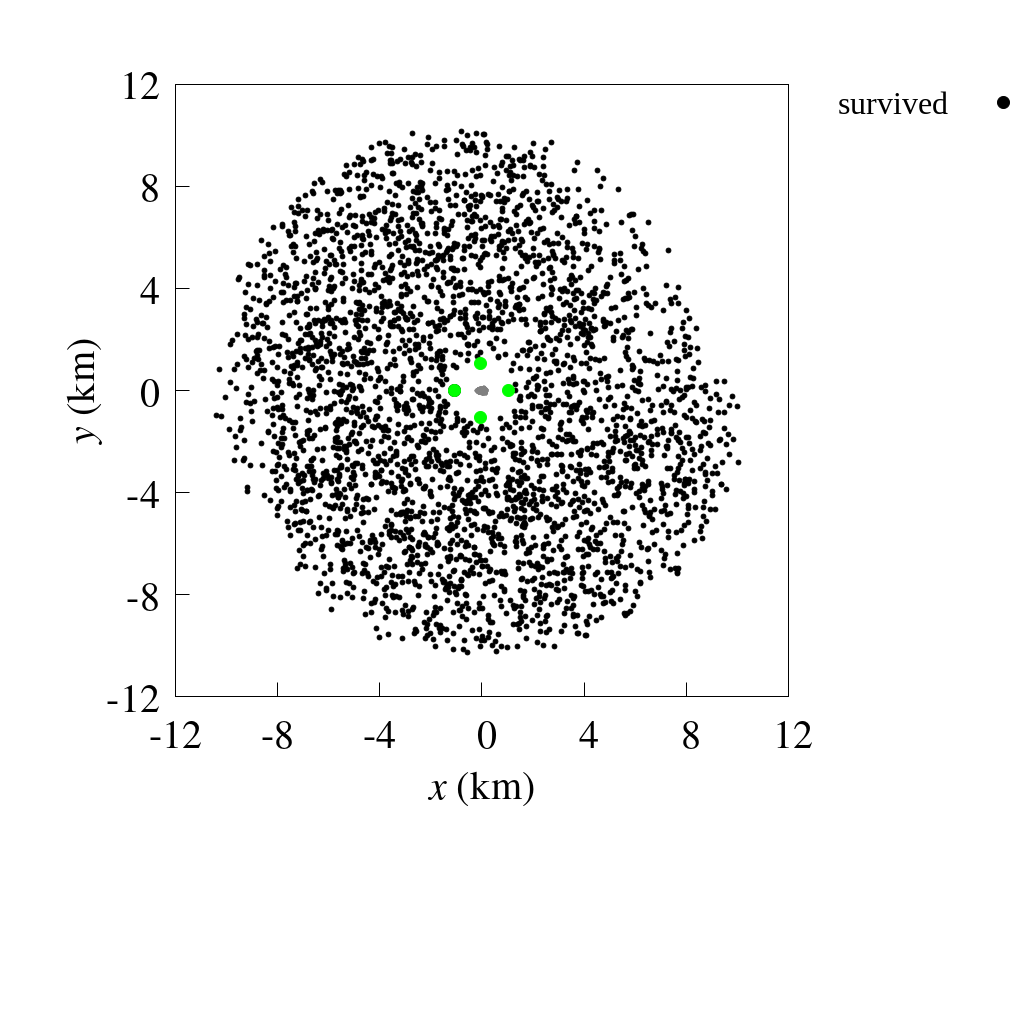}\label{gfig:24hT_129L}}
\subfloat[]{\includegraphics*[trim = 0mm 7cm 6.5cm 0mm,
width=0.66\columnwidth]{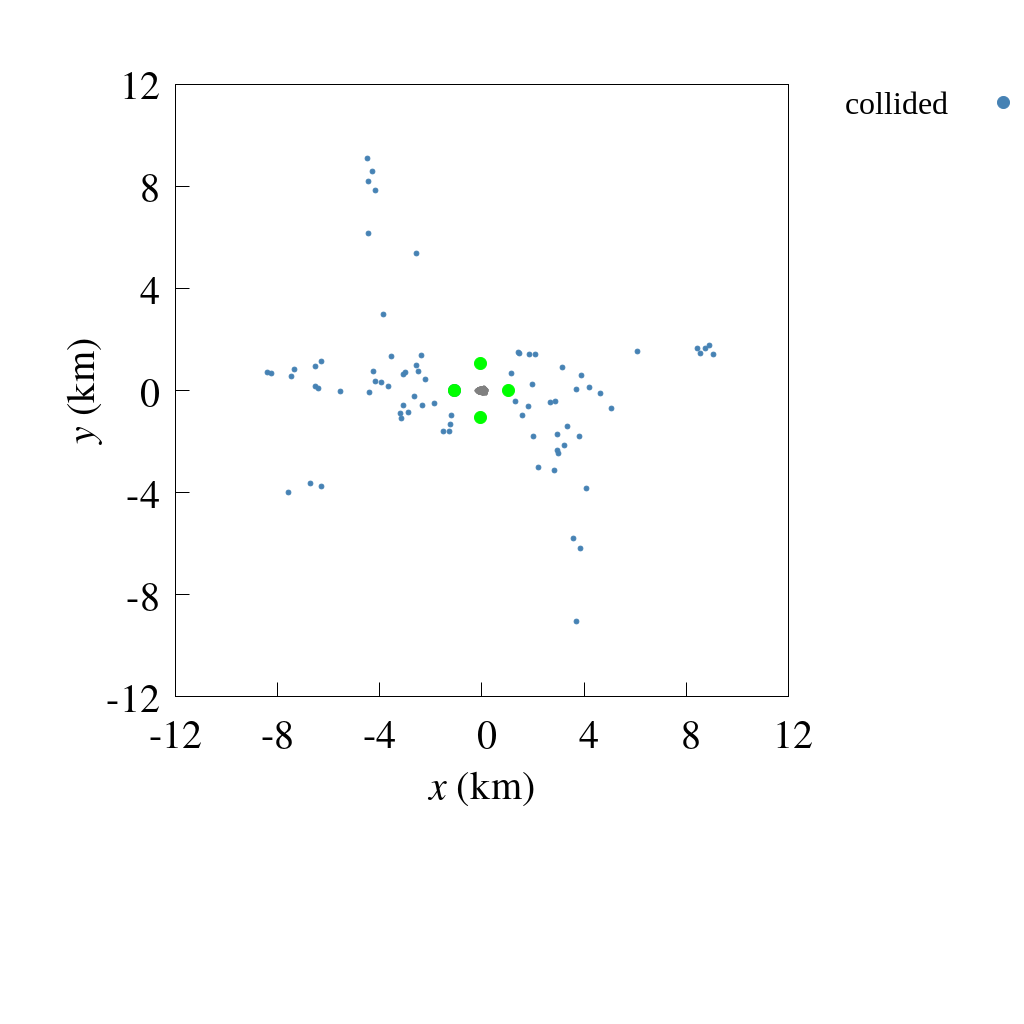}\label{hfig:24hT_22L}}
\subfloat[]{\includegraphics*[trim = 0mm 7cm 6.5cm 0mm, width=0.66\columnwidth]{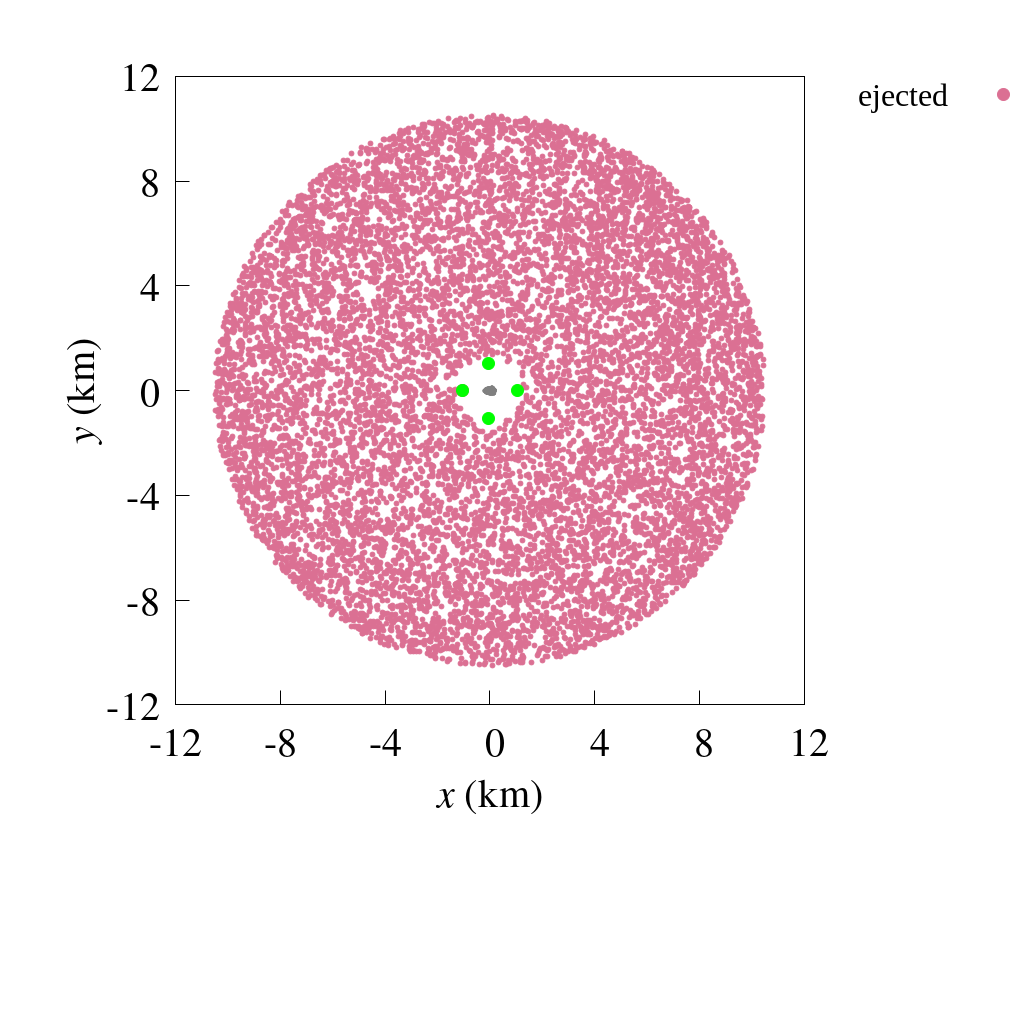}\label{ifig:24hT_35L}}
\end{center}
\caption{\label{fig:24hTL} Initial conditions of the 15 thousand particles of the expanded ring in the 2029 approach in the $xoy$ plane. The green dots represent the equilibrium points. The black dots represent the particles that survived after the 24 hours of integration and the blue and pink dots the particles that collided with the asteroid and were ejected, respectively. The letters (a)-(c) represent, respectively, the densities of 1.29 g$\cdot$cm$^{-3}$, 2.2 g$\cdot$cm$^{-3}$ and 3.5 g$\cdot$cm$^{-3}$.}
\end{figure*}

For the expanded radial distribution the number of ejections for the regular case (section \ref{regions}) was null, but for the encounter simulation was high (Fig. \ref{fig:24hTL}). About 75\% and 77\% of the particles were ejected for the smaller and larger model of density, respectively, being 1.3\% of this ejection by the energy criterion for the smaller and larger models. The percentage of survived particles was approximately 24.2\% for the model with the smaller density, while for the larger one was about 22.5\%. The number of collisions was small, about 0.8\% and 0.5\% for the smaller and larger model of density. 

The 2029 approach with the Earth shown to be significant in the nearby environment of Apophis for both disc cases. For the disc around the equilibrium points, the Earth's perturbation causes a high number of ejections, but also almost twice the value of collisions due to the gravitational field of this region being high. On the other hand for the expanded disc, the encounter causes a massive number of ejections and a smaller number of collisions and surviving particles.

\section{Final Comments}
\label{final}

In this study, we provided the exploration of the effects of the 2029 Earth’s encounter on the surface and nearby dynamics of the asteroid Apophis. Firstly, we briefly discuss the Apophis' shape model \citep{Pravec2014}, its physical properties as the density and size discordance \citep{BINZEL2009480, muller2014thermal, licandro2015canaricam, BROZOVIC2018115}, and the 2029 Earth’s encounter and its trajectory. Then, we define gravitational potential by the polyhedra method \citep{wernerscheeres1996} and the geopotential on the Apophis' surface \citep{Scheeres2012, Scheeres2016}.

We presented the slope angle and its implications on the body's surface. The slope computed on Apophis' surface was about 36$^{\circ}$ for the three density models with small variations among the models, a difference of only 0.35$^{\circ}$ between the smaller and larger density model. To comprehend the possible effect caused by the 2029 encounter, we calculated the variation of the slope angle produced by the Earth’s gravitational perturbation called $\Delta$slope. The $\Delta$slope was smaller than 4$^{\circ}$ and 2$^{\circ}$ for the density of 1.29 g$\cdot$cm$^{-3}$ and 3.5 g$\cdot$cm$^{-3}$, respectively. Those variations may cause migration of cohesionless particles since they can reach the value of the repose angle of geological materials (35$^{\circ}$-40$^{\circ}$) in some regions \citep{lambe1969, apollo1974, al2018review}. A variation of about 2$^{\circ}$ may generate a slow erosion process in those regions with high-slope \citep{ballouz2019surface}, but numerical simulations have shown that the perturbation of the Earth in the 2029 flyby may create just local landslides \citep{yu2014numerical}.

Next, we analysed the zero-velocity curves and computed the equilibrium points of the body. Apophis has four external equilibrium points and two of them are topologically classified as Centre-Centre-Centre for the three density models, implying that they are linearly stable points \citep{Jiang2014}. Thus, we performed a set of numerical simulations of a disc of 15,000 particles in the nearby environment of Apophis. The first set of simulations was performed considering just the gravitational perturbation of Apophis for a period of 24 hours and 30 years. For a period of 24 hours, the majority of the ring encompassing the equilibrium points is stable with just 0.78-1.4\% of collision and no ejection, while for the expanded disc all particles survived. For the period of 30 years, 96\% of the particles survived for the larger density model, 0.9\% collided and 3.1\% ejected in a region that has limited the influence of Apophis, creating an external disc of survived particles (Fig. \ref{cfig:30y_35}). The same phenomenon happens for the smaller density, however, 52\% of the particles survived, 4\% collided, and 44\% ejected. For the density of 1.29 g$\cdot$cm$^{-3}$, we found interesting regions of survived particles that are related to resonance regions and will be studied in future works.

We added the perturbation of the solar radiation pressure at the system and simulate the same initial conditions of the previous set of simulations, but now just for the period of 30 years. With those simulations, we identify that just centimetres particles survive for the whole period. For the density model of 2.2 g$\cdot$cm$^{-3}$ and 3.5 g$\cdot$cm$^{-3}$, we compute a significant amount of surviving particles with a radius of 5 cm and for the density model of 1.29 g$\cdot$cm$^{-3}$, just particles with radius of 15 cm survived, considering our set of discrete sizes of particles.

In addition, we performed simulations considering the 2029 encounter and the Earth's perturbation but removing the solar radiation pressure. Since the solar radiation pressure removes small particles, for this set of simulations we implicitly assume that all of the particles must have a centimetre size of about 5-15 cm. At the end of the 24 hours simulation for the disc encompassing the equilibrium points, we compute a survival percentage of 56\% for the larger density model and 59\% for the smaller one. However, for the expanded disc, the majority of the particles were ejected and just 22.5-24.2\% of particles survived. In general, the 2029 approach shows that the Earth perturbation is significant in the nearby environment of Apophis, but a considerable number of particles still remain at the end of the encounter.

All the analyses presented in this study, whose purpose was to identify the possible effects on Apophis' surface and nearby environment due to the 2029 flyby, may contribute to the 2029 observational campaign and also to validate our results.


\section*{Acknowledgements}

This study was financed in part by the Coordenação de Aperfeiçoamento de Pessoal de Nível Superior - Brasil (CAPES) - Finance Code 001, Fundação de Amparo à Pesquisa do Estado de São Paulo (FAPESP) - Proc. 2016/24561-0 and Proc. 2019/23963-5, Conselho Nacional de Desenvolvimento Científico e Tecnológico (CNPq) - Proc. 305210/2018-1.

\section*{ORCID iDs}
G. Valvano \orcidicon{0000-0002-7905-1788} \href{https://orcid.org/0000-0002-7905-1788}{https://orcid.org/0000-0002-7905-1788}\\
O. C. Winter \orcidicon{0000-0002-4901-3289} \href{https://orcid.org/0000-0002-4901-3289}{https://orcid.org/0000-0002-4901-3289}\\
R. Sfair \orcidicon{0000-0002-4939-013X} \href{https://orcid.org/0000-0002-4939-013X}{https://orcid.org/0000-0002-4939-013X}\\
R. Machado Oliveira \orcidicon{0000-0002-6875-0508} \href{https://orcid.org/0000-0002-6875-0508}{https://orcid.org/0000-0002-6875-0508}\\
G. Borderes-Motta \orcidicon{0000-0002-4680-8414} \href{https://orcid.org/0000-0002-4680-8414}{https://orcid.org/0000-0002-4680-8414}\\
T. S. Moura \orcidicon{0000-0002-3991-8738} \href{https://orcid.org/0000-0002-3991-8738}{https://orcid.org/0000-0002-3991-8738}\\

\section*{Data availability}
The data underlying this article will be shared on reasonable request to the corresponding authors.



\bibliographystyle{mnras}
\bibliography{bilbli} 


\newpage
\bsp	
\label{lastpage}
\end{document}